\documentclass[twocolumn,superscriptaddress,floatfix,preprintnumbers, nofootinbib,hyperref]{revtex4-1}
\usepackage{graphicx}% Include figure files
\usepackage{dcolumn}% Align table columns on decimal point
\usepackage{bm}% bold math
\PassOptionsToPackage{hyphens}{url}\usepackage{hyperref}% add hypertext capabilities
\usepackage[mathlines]{lineno}% Enable numbering of text and display math
\usepackage{float}
\usepackage{amsmath,amssymb,mathtools}					% packages for nicer use of mathematics
\usepackage{makecell}
\usepackage[version=4]{mhchem}
\usepackage{subfigure}
\usepackage{color}
\usepackage[normalem]{ulem}
\usepackage{textgreek}

\DeclareMathOperator{\GeV}{GeV}

\providecommand{\tabularnewline}{\\}
\newcommand{\gag}{g_{a\gamma}}

\begin{document}

\preprint{LAPTH-038/21}

%\title{%Constraining the diffuse supernova axion-like-particle background\\ with high-latitude \Fermi-LAT data
%\JJ{\JJC{Suggestion for title:} \\
%New \Fermi-LAT constraints on the diffuse axion-like particle background}}% Force line breaks with \\
\title{3D template-based  \Fermi-LAT constraints on the diffuse supernova \\ axion-like particle background$^{1}$
}
\footnotetext{Parts of this work have already been shown at ICRC 2021 and a short preview has appeared in the associated proceedings \cite{Eckner:2021Pe}.}
%\email[x]{Parts of this work have already been shown at ICRC 2021 and a short preview has appeared in the associated proceedings~\cite{Eckner:2021Pe}.}

\author{Francesca Calore}
\email{calore@lapth.cnrs.fr}
\affiliation{Univ.~Grenoble Alpes, USMB, CNRS, LAPTh, F-74000 Annecy, France}

\author{Pierluca Carenza}
\email{pierluca.carenza@fysik.su.se}
\affiliation{The Oskar Klein Centre, Department of Physics, Stockholm University, Stockholm 106 91, Sweden
}
\affiliation{Dipartimento Interateneo di Fisica ``Michelangelo Merlin'', Via Amendola 173, 70126 Bari, Italy.}
\affiliation{Istituto Nazionale di Fisica Nucleare - Sezione di Bari, Via Orabona 4, 70126 Bari, Italy.}

\author{Christopher Eckner}
 \email{eckner@lapth.cnrs.fr}
\affiliation{Univ.~Grenoble Alpes, USMB, CNRS, LAPTh, F-74000 Annecy, France}

\author{Tobias Fischer}
\email{tobias.fischer@uwr.edu.pl}
\affiliation{Institute of Theoretical Physics, University of Wroc{\l}aw, 50-204 Wroc{\l}aw, Poland}

\author{Maurizio Giannotti}
\email{mgiannotti@barry.edu}
\affiliation{Physical Sciences, Barry University, 11300 NE 2nd Ave., Miami Shores, FL 33161, USA}

\author{Joerg Jaeckel}
\email{jjaeckel@thphys.uni-heidelberg.de}
\affiliation{Institut f\"ur theoretische Physik, Universit\"at Heidelberg,
Philosophenweg 16, 69120 Heidelberg, Germany}

\author{Kei Kotake}
\affiliation{Faculty of Science, Department of Applied Physics \& Research
  Institute of  Stellar Explosive Phenomena, Fukuoka University, Fukuoka, Jonan 8-19-1, 814-0180, Japan}

\author{Takami Kuroda}
%\email{takami.kuroda@aei.mpg.de}
\affiliation{Max-Planck-Institut f{\"u}r Gravitationsphysik, Am M{\"u}hlenberg 1, D-14476 Potsdam-Golm, Germany}

\author{Alessandro Mirizzi}
\email{alessandro.mirizzi@ba.infn.it }
\affiliation{Dipartimento Interateneo di Fisica ``Michelangelo Merlin'', Via Amendola 173, 70126 Bari, Italy.}
\affiliation{Istituto Nazionale di Fisica Nucleare - Sezione di Bari, Via Orabona 4, 70126 Bari, Italy.}

\author{Francesco Sivo}
\email{francesco.sivo@ba.infn.it}
\affiliation{Dipartimento Interateneo di Fisica ``Michelangelo Merlin'', Via Amendola 173, 70126 Bari, Italy.}

\definecolor{darkblue}{rgb}{0, 0.1, 1.0}
\definecolor{newColor}{rgb}{1, 0, 1}

\newcommand{\JJ}[1]{{\color{darkblue}#1}}%blue
\newcommand{\JJC}[1]{{{\bf{\color{darkblue}#1}}}}%blue
\newcommand{\AM}[1]{{\color{red}#1}}%red
\newcommand{\MS}[1]{{\color{magenta}#1}}%magenta
\newcommand{\mau}[1]{{\color{newColor}#1}}%Maurizio
\newcommand{\mg}[1]{{\color{newColor}[\underline{\bf MG}:~{\bf #1}]}}%Maurizio
\newcommand{\deletemau}[1]{{\color{newColor}\sout{#1}}}%Maurizio
\newcommand{\PC}[1]{{\color{purple}#1}}%magenta
\definecolor{ballblue}{rgb}{0.13, 0.67,0.8}
\newcommand{\FC}[1]{{\color{ballblue}[FC:~#1]}}%Francesca comment
\newcommand{\fc}[1]{{\color{ballblue}#1}}%Francesca comment
\definecolor{redblick}{rgb}{0.796, 0.255, 0.329}
\newcommand{\TK}[1]{{\color{redblick}[TK:~#1]}}%Takami comment

\definecolor{forestgreen}{rgb}{0.13, 0.545, 0.13}
\newcommand{\CE}[1]{{\color{forestgreen}[CE:~#1]}}%Christopher's comments

\newcommand{\cec}[1]{{\color{forestgreen}#1}}%Christopher's comments
\newcommand{\FS}[1]{{\color{orange}[FS:~#1]}}%Francesco's comments

\newcommand{\NEW}[1]{{\color{red}#1}}%red
\newcommand{\NEWW}[1]{{\color{blue}#1}}%red

\newcommand{\Fermi}{{\it Fermi}}

\date{\today}% It is always \today, today,

\begin{abstract}
Axion-like particles (ALPs) may be abundantly produced in core-collapse (CC) supernovae (SNe), hence the cumulative signal from all past SN events can create a diffuse flux peaked at energies of about 25~MeV. 
We improve upon the modeling of the ALPs flux by including a set of CC SN models with different progenitor masses, as well as the effects of failed CC SNe -- which yield the formation of black holes instead of explosions. Relying on the coupling strength of ALPs to photons and the related Primakoff process, the diffuse SN ALP flux is converted into gamma rays while traversing the magnetic field of the Milky Way. The spatial morphology of this signal is expected to follow the shape of the Galactic magnetic field lines.
We make use of this via a template-based analysis that utilizes 12 years of \Fermi-LAT data in the energy range from 50 MeV to 500 GeV. In our benchmark case of the realization of astrophysical and cosmological parameters, we find an upper limit of $g_{a\gamma} \lesssim 3.76\times10^{-11}\;\mathrm{GeV}^{-1}$ at 95$\%$ confidence level for $m_a \ll 10^{-11}$ eV, while we find that systematic deviations from this benchmark scenario induce an uncertainty as large as about a factor of two.
Our result slightly improves the CAST  bound, while still being a factor of six (baseline scenario) weaker than the SN1987A gamma-ray burst limit.
\end{abstract}

\maketitle

%\tableofcontents

\section{Introduction}
\label{sec:intro} 
%\JJC{Titles are now included in bibliography... }
%\sout{The} {\bf A} gravitational {\bf core-}collapse \sout{of a} supernova (SN) \sout{would} {\bf can}  be a powerful source of axions and
%	 axion-like particles (ALPs)~\cite{Raffelt:2006cw,Fischer:2016cyd}.
Observing a supernova (SN) provides unique opportunities for fundamental physics. In particular, they are ideally suited to probe feebly interacting particles (cf.~\cite{Agrawal:2021dbo} for a recent review) with masses up to the $\sim 100\,{\rm MeV}$ range. Indeed, large numbers of such particles can be emitted in SN events~\cite{Raffelt:1996wa}.  An important and theoretically interesting instance of this are axions and axion-like particles (ALPs)~\cite{Raffelt:2006cw,Fischer:2016cyd}.
Indeed,	SN~1987A has significantly strengthened  astrophysical axion bounds in a region of the yet-incompletely known ALP parameter space, complementary to the one 
	probed by the Sun and the globular clusters~\cite{Raffelt:2006cw,Brockway:1996yr,Grifols:1996id,Carenza:2019pxu,Lucente:2020whw}.
	In the minimal scenario in which ALPs are coupled only with photons, the main channel for their emissivity in the SN core is the Primakoff
	process, leading to an ALP flux peaked at energies of about 25 MeV. 
	%\fc{\sout{In case of ultralight ALPs ($m_a \ll 10^{-10}$~eV), for a Galactic SN this flux can be converted into gamma-rays in the Milky Way magnetic field, leading to an observable gamma-ray burst in coincidence with the SN explosion~\cite{Brockway:1996yr,Grifols:1996id}.} 
	Conversion of these ALPs into gamma rays in the Milky Way magnetic field can lead to an observable gamma-ray burst in coincidence with the
	SN explosion~\cite{Brockway:1996yr,Grifols:1996id}. 
 At the time of the SN 1987A, the  Gamma-Ray Spectrometer (GRS) on the Solar Maximum Mission (SMM) 
 %\sout{had the SN in its field of view and} 
 observed no gamma-ray signal at the time of the SN explosion, this made possible to constrain the photon-ALP coupling early on (see \cite{Brockway:1996yr, Grifols:1996id} for a detailed discussion). In a more refined, recent analysis, this upper limit is stated as
	$g_{a\gamma} \lesssim 5.3 \times 10^{-12}\,\textrm{GeV}^{-1}$ for $m_a <  4 \times10^{-10}$~eV~\cite{Payez:2014xsa}. X-ray observations of super star clusters in the vicinity of the Milky Way's Galactic center can further strengthen this bound to $g_{a\gamma} \lesssim 3.6 \times 10^{-12}\,\textrm{GeV}^{-1}$ for $m_a <  5 \times10^{-11}$~eV at $95\%$ confidence level (CL) \cite{Dessert:2020lil}.
%	\fc{\sout{Forecasts in for the case of a Galactic SN explosion observed by \Fermi~Large Telescope Array (LAT) are really promising, allowing us} 
A future Galactic SN explosion in the field of view of the Large Area Telescope (LAT) aboard the \Fermi~satellite would allow us 
	to constrain $g_{a\gamma} \lesssim 2.0 \times 10^{-13}\,\textrm{GeV}^{-1}$ for $m_a <  10^{-9}$~eV~\cite{Meyer:2016wrm}. 
	Furthermore, a search for gamma-ray bursts from extragalactic SNe with \Fermi-LAT has yielded the limit 
	$g_{a\gamma} \lesssim 2.6 \times 10^{-11}\,\textrm{GeV}^{-1}$ for $m_a <  3 \times 10^{-10}$~eV~\cite{Meyer:2020vzy}
		(see also~\cite{Meyer:2021ont,Crnogorcevic:2021wyj}).
%	\fc{\sout{under the assumption of at least one SN occurring in the detector field of view~\cite{Meyer:2020vzy} (see also~\cite{Meyer:2021ont,Crnogorcevic:2021wyj}).}}
	
%	\JJC{Maybe put the following paragraph at the beginning and shorten things before???}
%\fc{\sout{	 A next Galactic SN would be the Holy Grail for ALP astrophysics. However, these events are rare and one has to be lucky to detect one of them in the lifetime of \Fermi-LAT. Therefore, waiting for the next Galactic explosion one can take advantage of a guaranteed SN ALP flux, namely the Diffuse SN ALP Background (DSNALPB) coming from all past core-collapse SN in the Universe~\cite{Raffelt:2011ft}.}}
	 While a single SN event is rare \citep{diel06} and must fall into the detector field of view to be observed, there exists a guaranteed contribution to the gamma-ray diffuse flux which originates from ALPs emitted by all past SNe in the Universe~\cite{Raffelt:2011ft}.
	 This Diffuse SN ALP Background (DSNALPB), 
	 despite being fainter
	 than the Galactic one, is within the reach of the \Fermi-LAT experiment. %\fc{\sout{Therefore, it is worthwhile to try to get the most from it concerning ALPs. }}
	In Ref.~\cite{Calore:2020tjw}, henceforth called `Paper I', some of us used published \Fermi-LAT observations of the gamma-ray isotropic diffuse background to set a bound $g_{a\gamma} \lesssim 5.0 \times 10^{-11}\,\textrm{GeV}^{-1}$ for $m_a <  10^{-11}$~eV. 
	However, this analysis does not completely acknowledge some technical issues behind the derivation of the isotropic gamma-ray background, which may impact the reliability of the stated upper bound on $g_{a\gamma}$. This component of the gamma-ray sky is obtained in connection with a particular model of the diffuse gamma-ray flux from the Milky Way and evaluated in a particular region of interest (ROI). Both the dependence on the diffuse model and the dependence on the selected sky region introduce unknowns in the upper bound estimate that cannot be cast into an uncertainty on the derived value because it is not known if the initial choices made by the \Fermi-LAT collaboration create artificially strong or weak limits. Hence, we deem it warranted to put the analysis of the DSNALPB on solid statistical foundations by creating a complete \Fermi-LAT data analysis pipeline which takes into account all the experience that has been gained over the long run of the LAT.
	%\fc{\sout{using the  recent measurements of the diffuse gamma-ray flux observed by the \Fermi-LAT.}}
	
	% present work
	In the present work, %\footnote{\cec{Parts of this work have already been shown at ICRC 2021 and a \JJ{short preview} has appeared in the associated proceedings \cite{Eckner:2021Pe}.}}, 
	we improve upon the previous 
	analysis presented in Paper I in two ways.
	First, we present a more refined model of the SNe ALPs flux. It is indeed well known that the production of ALPs in a SN event depends on the progenitor mass. In Paper I, however, it was assumed that all past SNe are represented by a 18 $M_{\odot}$ progenitor model. Here, instead, we consider
	different CC SN models with masses ranging between 8.8 and 70 $M_{\odot}$, accounting also for  the contribution 
	due to failed\footnote{In the supernova models considered here, ``failed'' supernova is defined by a model with BH formation or without a shock revival during the numerical simulation. We will see the effect this has on the ALP production momentarily.} core-collapse (CC) SN explosions. This allows us to determine with better accuracy a possible range
	of variability of the DSNALPB, and, in turn, of the expected gamma-ray flux.
	Secondly, we try to exploit the full potential of \Fermi-LAT data in searching for this type of signal, by including information on the expected spatial structure of the signal in the gamma-ray data analysis. Paper I indeed sets limits on ALPs solely making use of the spectral energy distribution of the data. 
	On the other hand, template-based analyses -- see e.g.~\cite{1996ApJ...461..396M} for an early application in the context of EGRET data or a more recent example of an analysis of \emph{Fermi}-LAT data \cite{2010ApJ...724.1044S} that led to the discovery of the so-called \emph{Fermi Bubbles} -- exploit both spectral and spatial properties of gamma-ray data to constrain physics models. 
	This gamma-ray fitting technique has proven to be particularly successful in testing the hypothesis of weakly interacting massive particles shining in gamma rays at GeV - TeV energies (see, for instance, \cite{2012ApJ...761...91A, TheFermi-LAT:2017vmf, Storm:2017arh, DiMauro:2019frs, DiMauro:2021raz}). 
	However, to our knowledge, it was never applied to the search of an ALPs signal, albeit it presents specific spatial features, as we discuss below.
	We therefore perform a template-based analysis to constrain the ALP parameter space via the spatial structure of the DSNALPB induced diffuse gamma-ray flux using 12 years of \Fermi-LAT data in the energy range from 50 MeV to 500 GeV.
	
%	The plan of our work is as follows.  
%Signal: axion produced in CCSNe in the Universe and converted into photons into Galactic magnetic field.

%Goal: set bounds using \Fermi-LAT data with a template fitting approach in order to fully exploit the spatial features of the signal. 

%Novelties: 1. updated calculation of ALPs production in CC cosmological SNe, considering different mass progenitors for the SNe and inclusing the possibility of failed SNe; 2. Template fit to \Fermi-LAT data to maximize the sensitivity to the ALPs signal by exploiting its spatial distribution. 
The paper is organized as follows.
In Sec.~\ref{sec:SNmodels}, we illustrate
the CC SN models based on state-of-the-art hydrodynamical simulations.  
In Sec.~\ref{sec:alpflux}, we present our updated calculation of the ALPs production flux in SNe and induced   gamma-ray flux from the DSNALPB.
In Sec.~\ref{sec:data-analysis}, we sketch the analysis framework: data selection and preparation, and template fitting method of \Fermi-LAT data.
We discuss our results in Sec.~\ref{sec:results}.
We discuss systematic uncertainties and their impact on the ALPs upper limits in Sec.~\ref{sec:discussion}, and conclude in Sec.~\ref{sec:conclusion}.
%\NEW{\sout{It follows an Appendix where it is characterized the effect of the gravitational energy-redshift on the SN ALP spectra due to the strong gravitational field of the proto-neutron star}}\JJ{\sout{that was not included in the previous analysis}}\footnote{\JJ{\sout{We would like to thank the anonymous referee for making us aware of this effect}}.}
%}
Two final appendices are devoted to more technical issues. 
In Appendix
 
\ref{app:A} we characterize some details concerning the calculation of the SN ALP spectrum, namely the effect of the presence of alpha particles in the SN core and   the effects of the gravitational energy-redshift  due to the strong gravitational field of the proto-neutron star, which were overlooked in previous analyses.\footnote{We are grateful to the anonymous referee for bringing the relevance of these effects to our attention.}
In Appendix \ref{app:B}, we present more details on the systematic uncertainty on the DSNALPB upper limits of cosmological and astrophysical origin.

\section{Core-collapse supernova models}
\label{sec:SNmodels}
In order to provide reliable constraints on the DSNALPB, it is essential to cover a representative, wide range of SN models, which are based on state-of-the-art simulations. The present work discusses SN simulations which are based on general relativistic neutrino radiation hydrodynamics featuring three-flavor neutrino transport, both in spherical symmetry~\cite{Mezzacappa:1993gn,Liebendoerfer:2004,Fischer:2016,Fischer:2020} with accurate Boltzmann neutrino transport, and in axial symmetry with a multi-energy neutrino transport method~\cite{Kuroda:2021}. These simulations implement a complete set of weak interactions~\cite{Kotake:2018}, and a multi-purpose microscopic nuclear matter equation of state (EOS)~\cite{Lattimer:1991nc,Shen:1998gg,Hempel:2009mc,Hempel:2012,Steiner:2013}. 

\begin{figure*}[t!]
\vspace{0.cm}
\includegraphics[width=0.975\columnwidth]{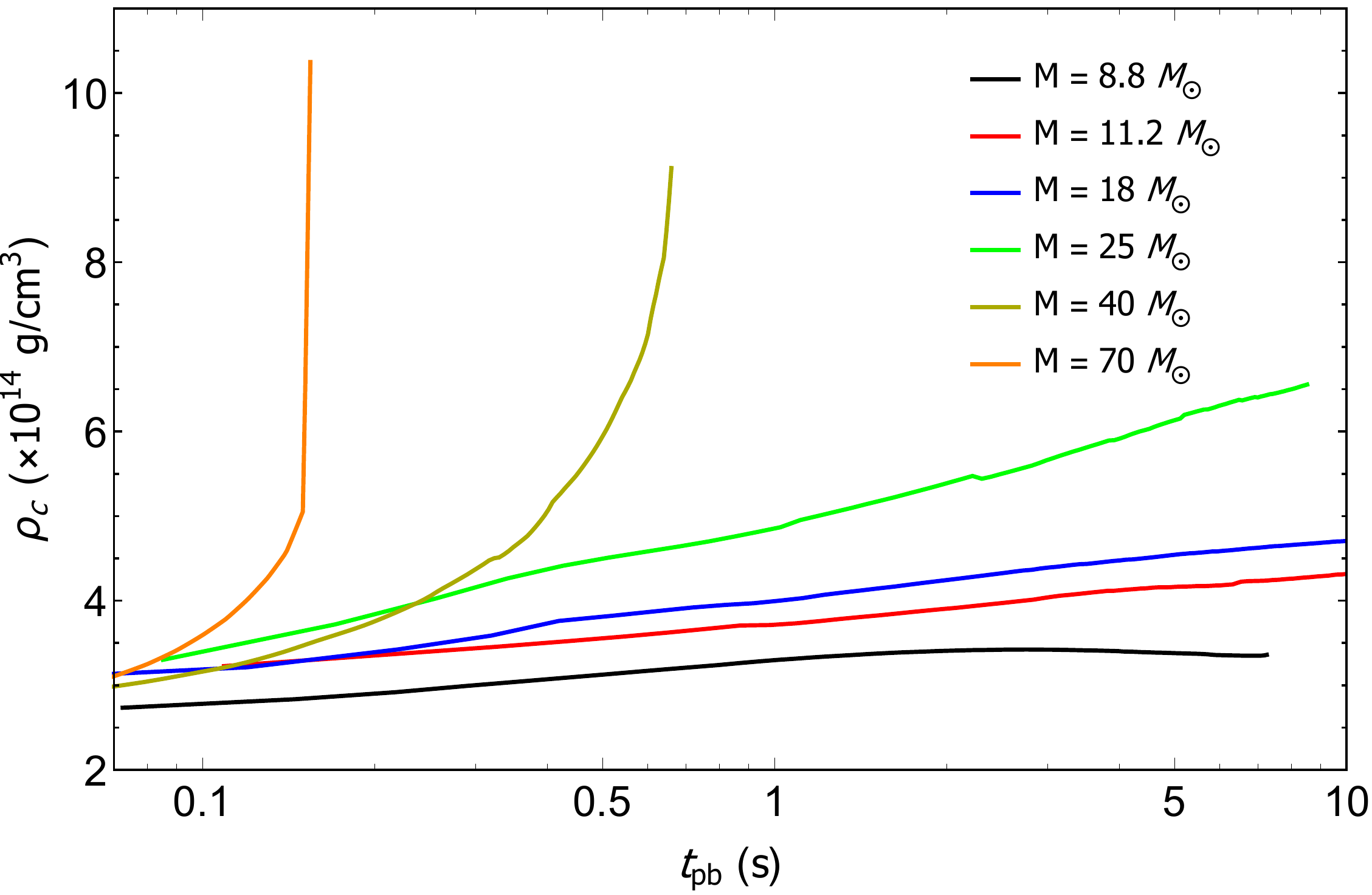}
\hfill
\includegraphics[width=0.975\columnwidth]{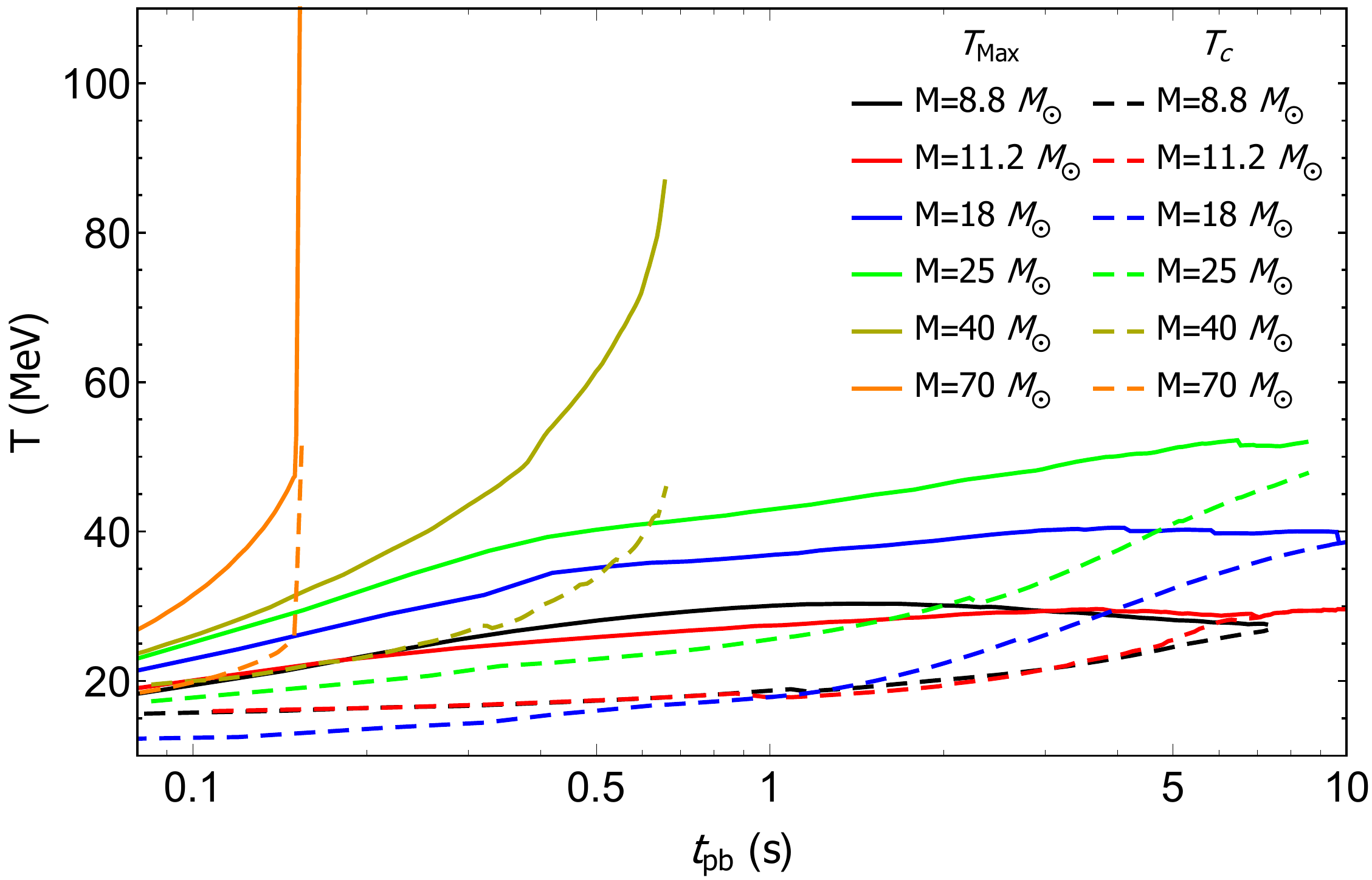}
\caption{PNS evolution during the deleptonization phase for the SN explosion models launched from different progenitors with ZAMS masses of 8.8, 11.2, 18.0 and $25~M_\odot$. {\em Left panel}:~central density, $\rho_{\rm centre}$. {\em Right panel}:~central and maximum temperatures, $T_{\rm centre}$ (dashed lines) and $T_{\rm Max}$ (solid lines). \textcolor{red}
%\JJC{Would it make sense to indicate that the flux essentially terminates (for the heavy ones). Or perhaps mention it in the caption???}
%\TK{I have added a sentence in the main text (below in this page), in which I try to mention that the BH formation occurs immediately and thus ALPs emission is expected to cease. Does it make sense?}
}
\label{fig:rho_T}
\end{figure*}

In what follows, we distinguish successful core-collapse SN explosions of different types of progenitors. We consider the low-mass oxygen-neon-magnesium core progenitor with zero-age main sequence (ZAMS) mass of $8.8~M_\odot$~\cite{Nomoto:1987}. They belong to the class of electron-capture SN~\cite{Kitaura:2006}, which yield neutrino-driven SN explosions even in spherical symmetry. The SN simulations discussed here were reported in Ref.~\cite{Fischer:2009af}, based on the nuclear EOS of Ref.~\cite{Shen:1998gg}. The simulations include all SN phases, i.e. stellar core collapse, core bounce\footnote{We define the point in time of the core bounce when the maximum central density is reached at the end of the stellar core collapse, which coincides with the time of shock breakout.} with the formation of the bounce shock, the subsequent SN post-bounce mass accretion phase including the explosion onset with the revival of the stalled bounce shock and finally the long-term deleptonization phase of the compact hot and dense central remnant proto-neutron star (PNS). The latter SN phase is of particular importance for the emission of axions. The remnant of this electron-capture SN explosion is a low-mass neutron star with a baryon mass of about $1.37~M_\odot$. The corresponding PNS deleptonization features a nearly constant central density of $\rho_{\rm central}\simeq 3.5\times 10^{14}$~g~cm$^{-3}$ as well as central temperature decreasing from $T_{\rm Max}\simeq30$~MeV to 25~MeV during the PNS deleptonization up to about 7.5~s post bounce, as illustrated in Fig.~\ref{fig:rho_T}. In addition to the decreasing, central temperature, we show the maximum temperature evolution in Fig.~\ref{fig:rho_T}, which rises moderately from $T_{\rm core}\simeq20$~MeV to about 25~MeV. 

As an example of a low-mass iron-core progenitor we consider the example with ZAMS mass of $11.2~M_\odot$ from the stellar evolution series of Ref.~\cite{Woosley:2002zz}. In contrast to electron-capture SN, which are characterized by a short post-bounce mass accretion period on the order of only few tenths of a second before the onset of the explosion, more massive iron-core progenitors suffer from extended post-bounce mass accretion periods, which fail to yield neutrino-driven explosions in self-consistent spherically symmetric simulations. Nevertheless, in order to obtain explosions, the neutrino heating and cooling rates have been enhanced artificially in Ref.~\cite{Fischer:2009af}, which lead to the successful revival of the stalled bounce shock. It results in the SN explosion onset\footnote{In all these SN simulations, the
onset of the explosions is defined when the expanding shock wave reaches a radius of about 1000~km.} about 300~ms after core bounce, for this progenitor star of $11.2~M_\odot$. The subsequent evolution of the central density of $\rho_{\rm centre}\simeq 4\times 10^{14}$~g~cm$^{-3}$ as well as the central and maximum temperatures is illustrated in Fig.~\ref{fig:rho_T}. The latter differ only marginally from those of the $8.8~M_\odot$ model. 

Two more massive iron-core progenitors are included here, with ZAMS masses of $18.0~M_\odot$ and $25~M_\odot$, which are evolved in a similar fashion as the $11.2~M_\odot$ model leading to neutrino-driven SN explosions on the order of several hundreds of milliseconds after core bounce. However, the remnant PNSs are more massive and hence feature a higher central density as well as higher central and maximum temperatures than the 8.8 and $11.2~M_\odot$ models (see Fig.~\ref{fig:rho_T}). In particular the $25~M_\odot$ simulation reaches maximum temperatures at the PNS interior which reach as high as 50~MeV during the PNS deleptonization phase. This aspect is important for the axion emission since the axion emissivity has a strong temperature dependence. 

In addition to the successful CC SN explosion models, we consider two examples with ZAMS masses of $40$ and $70~M_\odot$ belonging to the failed  CC SN branch which yield the formation of black holes instead~\cite{Baumgarte:1996iu,Sumiyoshi:2006id,Fischer:2009,OConnor:2011}. In such case the mass accretion onto the bounce shock, in combination with the failed shock revival, leads to the continuous growth of the enclosed mass of the PNS until it exceeds the maximum mass given by the nuclear EOS, on a timescale of several hundreds of milliseconds up to one second post bounce. If no phase transition is considered~\cite{Fischer:2018,Fischer:2021}, the PNS collapses eventually and a black hole forms. The data for the SN simulation of the $40~M_\odot$ progenitor discussed in the following are taken from Ref.~\cite{Fischer:2009} based on the nuclear EOS of Ref.~\cite{Lattimer:1991nc}. It results in black hole formation at about 450~ms post bounce with an enclosed PNS mass of about $2.5~M_\odot$. The most massive progenitor model considered of $70~M_\odot$, belongs to the class of zero-metallicity stars~\citep{Takahashi14} for which a black hole forms within a few hundred milliseconds after core bounce~\citep{KurodaT18,Shibagaki21,Kuroda:2021}. This model has been evolved in axially symmetric simulations. 
%\TK{
Although the original SN simulation~\citep{Kuroda:2021} takes into account the effect of strong phase transition from nuclear matter to the quark-gluon plasma at high baryon density, the central quark core immediately collapses into a BH within $\sim1$~ms after its formation.
Therefore its influence on ALP emission is expected to be minor. Furthermore, as the central high temperature region is swallowed by the BH, most of the ALP emission is expected to cease abruptly once the BH formation occurs, as indicated by ending the lines in Fig.~\ref{fig:rho_T}.
%}
The corresponding baryonic PNS mass at the onset of the PNS collapse is estimated to be $\sim 2.6~M_\odot$. In comparison to the other SN explosion models, with ZAMS masses of $8.8-25~M_\odot$, the failed SN branch yields significantly higher central densities as well as core temperatures. The latter reaches shortly before black hole formation up to $\rho_{\rm centre}\geq 10^{15}$~g~cm$^{-3}$ and $T_{\rm Max}\geq 100$~MeV.

Having a set of characteristic supernovae, we will then do a simple interpolation between them, as will be described below.
%
% progenitor star has relatively higher core entropy(temperature) of $\sim1.6$ $k_{\rm B}$ baryon$^{-1}$ ($\sim9.7\times10^9$ K) at the onset of iron core collapse. At $\sim155$ ms after bounce, the central lapse function decreases below $\lesssim0.01$ and we observe a formation of the apparent horizon.	

\section{DSNALPB and gamma-ray flux}\label{sec:alpflux}
\subsection{ALPs emission from SNe}
	
We consider  a minimal scenario in which ALPs have only a two-photon coupling, characterized  by the 
Lagrangian~\cite{Raffelt:1987im}
	%....................................................................
	\begin{equation}
		\label{mr}
		{\cal L}_{a\gamma}=-\frac{1}{4} \, \gag
		F_{\mu\nu}\tilde{F}^{\mu\nu}a=\gag \, {\bf E}\cdot{\bf B}\,a~.
	\end{equation}
	%.....................................................................................
	Through this interactions ALPs may be produced in stellar plasma 
	primarily via the Primakoff process~\cite{Raffelt:1985nk}. In such a process thermal photons are converted into ALPs in the electrostatic field of ions, electrons and protons.
	We calculate the ALP production rate (per volume) in a SN core via Primakoff process closely following~\cite{Payez:2014xsa}, which finds 
	\begin{eqnarray}
		\dfrac{d \dot n_a}{dE}&=&
		\frac{g_{a\gamma}^{2}\xi^2\, T^3\,E^2}{8\pi^3\, \left( e^{E/T}-1\right) } \nonumber \\
		& &\left[ \left( 1+\dfrac{\xi^2 T^2}{E^2}\right)  \ln(1+E^2/\xi^2T^2) -1 \right] \,.
		\label{eq:axprod}
	\end{eqnarray}
	Here, $E$ is the photon energy
	measured by a local observer at the emission radius, $T$ the temperature and $\xi^2={\kappa^2}/{4T^2}$ with $\kappa$ the inverse Debye screening length, describing 
	the finite range of the electric field surrounding charged particles in the plasma.  
	The total ALP production rate per unit energy is obtained integrating Eq.~\eqref{eq:axprod} over the SN volume.
	Details on the calculation of the SN ALP spectrum are provided in Appendix \ref{app:A}. In particular, we discuss the enhancement of the ALP flux associated with the presence of  alpha particles in the SN core. Furthermore
		we also show that the strong gravitational field of a proto-neutron star can modify the ALP emissivity in the SN core via three General Relativistic effects: time dilation, trajectory bending and energy redshift.
		As explained in Appendix~\ref{app:GR}, 
		the trajectory bending has no effect on the time-integrated diffuse ALP background we aim to calculate. Therefore, we will focus only on energy-redshift and time-dilation here.

	Assuming $m_a \ll T$,  the  ALP 
	fluence
	is given, with excellent precision, by the analytical expression~\cite{Payez:2014xsa}
		%%%%%%%	
	\begin{equation}
		\frac{dN_a}{dE} = C \left(\frac{g_{a\gamma}}{10^{-11}\textrm{GeV}^{-1}}\right)^{2}
		\left(\frac{E}{E_0}\right)^\beta \exp\left( -\frac{(\beta + 1) E}{E_0}\right) \,,
		\label{eq:time-int-spec}
	\end{equation}
	%%%%%%%%%
	where the values of the 
	parameters $C$, $E_0$, and $\beta$  for the SN models with different progenitors  are given in Table~\ref{tab:spectrumfitting}.
	The spectrum described in Eq.~\eqref{eq:time-int-spec} is a typical quasi-thermal spectrum, with mean energy $E_0$ and index $\beta$ (in particular, $\beta=2$ corresponds to a perfectly thermal spectrum of ultrarelativistic particles).

	%%%%%%%%%%%%%%%%%%%%%%%%%%%%%%%
	\begin{table}[!t]
		\caption{Fitting parameters for the SN ALP spectrum, Eq.~\eqref{eq:time-int-spec}, from the Primakoff process for different SN progenitors.}
% 		\NEW{Fitting parameters for the SN ALP spectrum, Eq.~\eqref{eq:time-int-spec}, from the Primakoff process for different SN progenitors estimated for \sout{$g_{a\gamma}=10^{-11}\,\textrm{GeV}^{-1}$ and} 
% 		$m_a \ll 10^{-11}$~eV. \mg{The parameters we give here don't depend on $g_{a\gamma}$. 
% 		The coupling is factored out.
% 		I am also confused about the requirement that $m_a \ll 10^{-11}$~eV. In Eq. (3) we simply have the ALP spectrum for which it is sufficient to require $m_a \ll T$. }
% 		\JJC{Agreed, I would think that we only need $m_{a}\ll 1\,{\rm MeV}$. For the coupling we also only require that the coupling is sufficiently small such that backreaction can be neglected. This also goes for several of the following figures etc.}}
		
		\begin{center}
			\begin{tabular}{lccc}
				\hline
				SN progenitor & $ $
				$C$ [$\times 10^{50} \,\  {\rm MeV}^{-1}$] \,\ 
				& $E_0$ [MeV] &$\beta$ \\
				\hline
				\hline
				8.8 $M_{\odot}$ & $3.76  \,\     $ & 76.44 &2.59 \\
				11.2 $M_{\odot}$ & $7.09  \,\     $ & 75.70 &2.80 \\
				18 $M_{\odot}$ & $23.0   \,\    $ & 91.61 &2.43 \\
				25 $M_{\odot}$ & $28.1   \,\    $ & 105.5 &2.30 \\
				40 $M_{\odot}$ & $2.48   \,\   $ & 112.7 &1.92 \\
				70 $M_{\odot}$ & $0.391  \,\     $ & 30.44 & 0.785 \\
				\hline
			\end{tabular}
			\label{tab:spectrumfitting}
		\end{center}
	\end{table}
	%%%%%%%%%%%%%%%%%%%%%%%%%%%%%%%%
	
	%%%%%%%%%%%%%%%%%%%
	\begin{figure}[t!]
		\vspace{0.cm}
		\includegraphics[width=0.95\columnwidth]{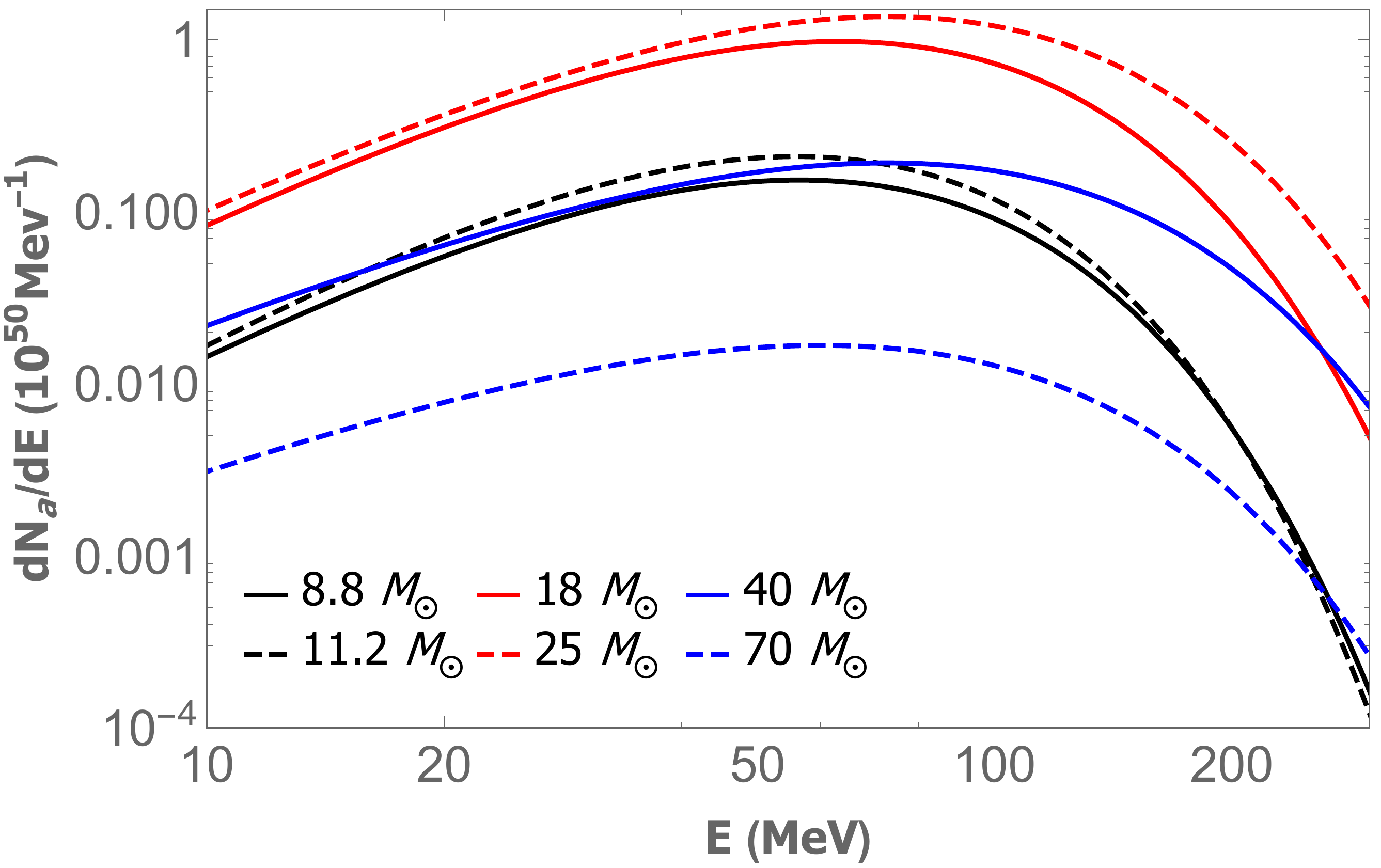}
		\caption{Produced SN ALP number as a function of energy for different SN progenitor mass.
			We assume 
			%\sout{$m_a \ll 10^{-11}$~eV} 
			$g_{a\gamma}=10^{-11}\GeV^{-1}$.
			}
		\label{fig:dNadE}
	\end{figure}
	%%%%%%%%%%%
	
	In Fig.~\ref{fig:dNadE}, we represent the SN ALP spectra from different progenitors. We realize that for the 
	successful CC SN explosions the  average energy $E_0$ increases monotonically with the progenitor mass, as well 
	as the peak of the spectrum. For the failed CC SN explosions, since the emitted flux is integrated over a shorter time window,
	the flux is suppressed with respect to the previous models.
	%\mg{I removed "Conversely" since to me the second sentence is not related to the first. It is just a separate statement. }

\begin{figure}[t!]
		\subfigure{
			\vspace{0.0cm}
			\includegraphics[width=0.95\columnwidth]{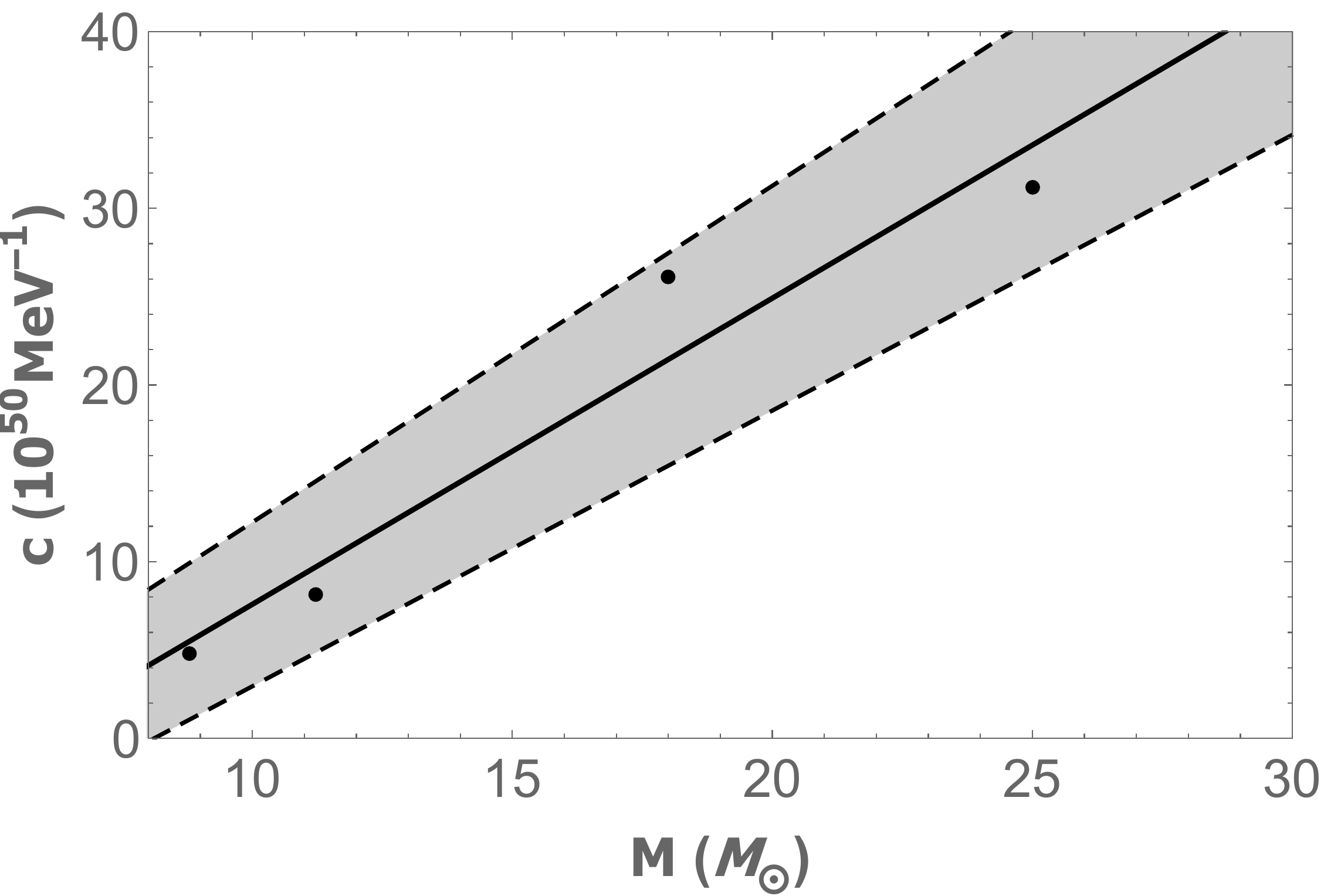}
		}
		\subfigure{
			\vspace{0.0cm}
			\includegraphics[width=0.95\columnwidth]{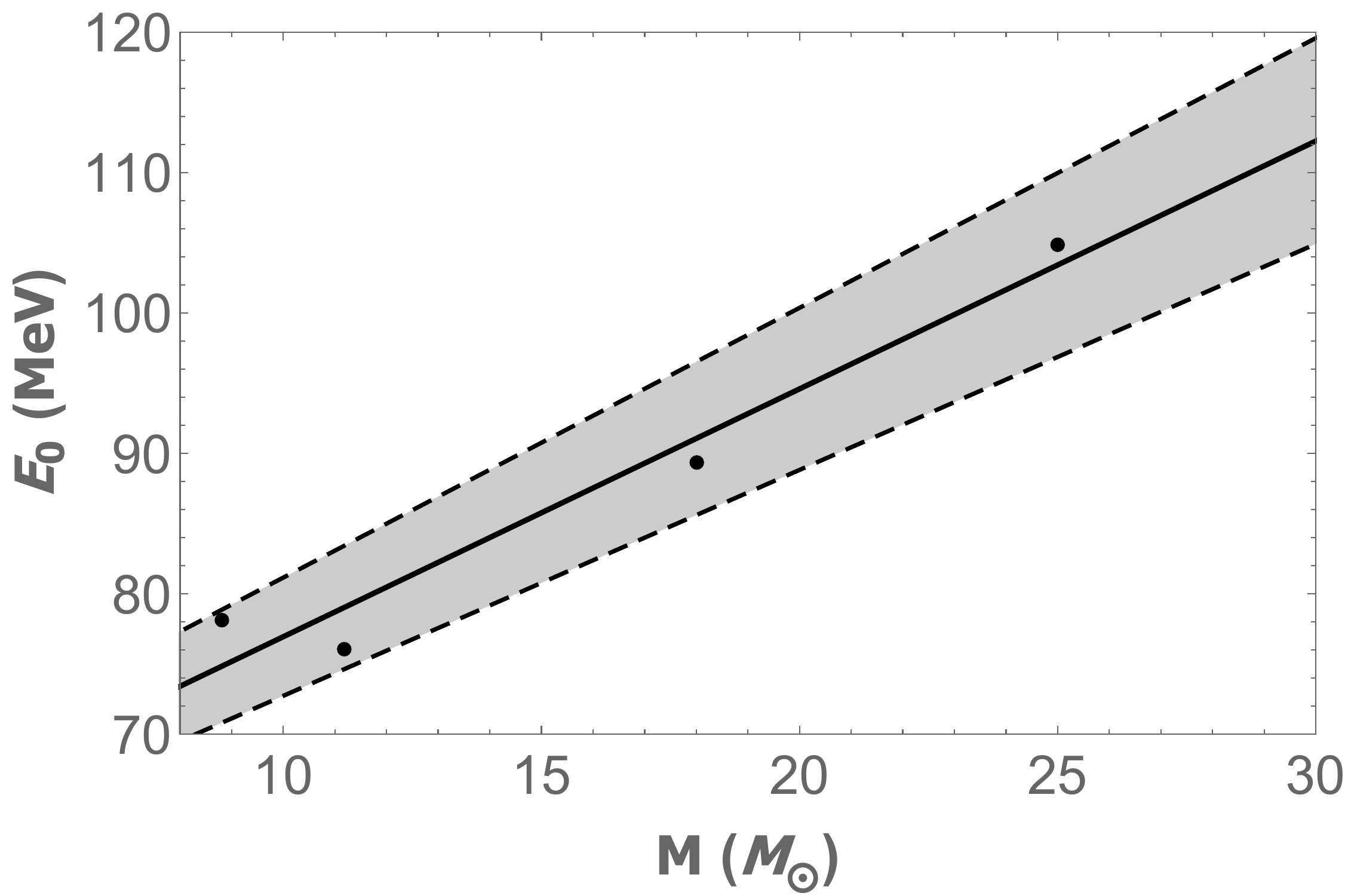}
		}
		\subfigure{
			\vspace{0.0cm}
			\includegraphics[width=0.95\columnwidth]{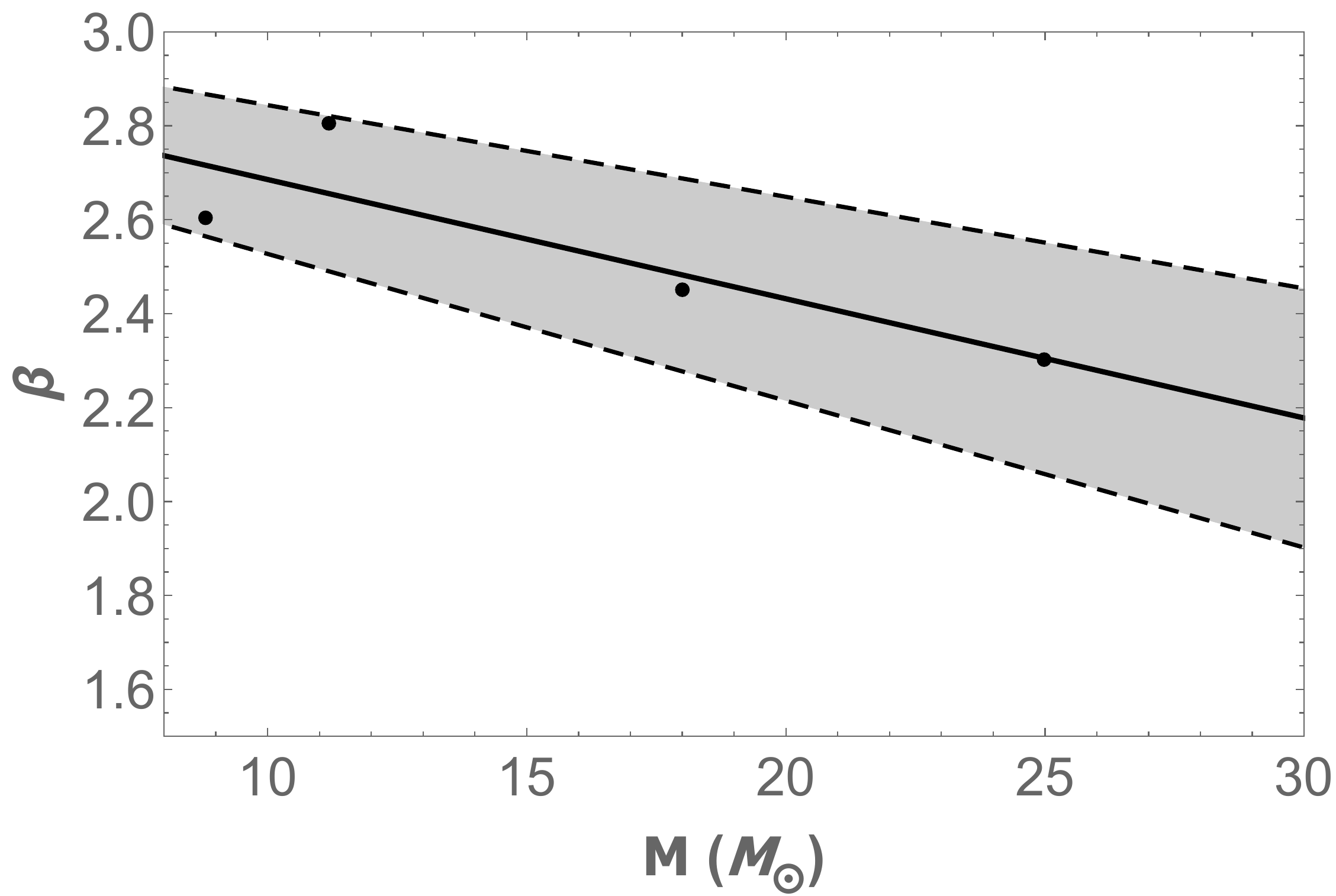}
		}
		\caption{Variation of spectral parameters of Eq.~(\ref{eq:time-int-spec}) as a function of the SN progenitor mass. We assume 
			$m_a \ll 10^{-11}$~eV and
			$g_{a\gamma}=10^{-11}\GeV^{-1}$.
		}
		\label{fig:fit}
	\end{figure}

	For further purposes related to the calculation of the DSNALPB it is useful to determine the variation of the 
	spectral coefficients $C$, $E_0$, and $\beta$ as a function of the SN progenitor mass.
	Given the sparseness of the data we assume a linear behaviour in the range $[8;30] M_{\odot}$,
	as shown in Fig.~\ref{fig:fit}. The functional expressions are the following ones
	%%%%%%%%%%%%
	\begin{eqnarray}
		\frac{C(M)}{10^{50} \,\ \textrm{MeV}^{-1}} &=& (1.73\pm 0.172)  \frac{M}{M_{\odot} } -9.74\pm 2.92   \,\ , \nonumber \\
	 	\frac{E_{0}(M)}{\textrm{MeV}} &=& (1.77\pm 0.156)  \frac{M}{M_{\odot} }+ 59.3\pm 2.65  \,\ , \nonumber \\
		\beta (M) &=& (-0.0254\pm 0.00587) \frac{M}{M_{\odot} } + 2.94\pm 0.0997  \,\ , \nonumber  \\
		\label{eq:parameters}
	\end{eqnarray}
		where the quoted errors represent the standard mean-square uncertainties associated with the linear regression, and are taken into account into the final evaluation of the uncertainty on the bound.\\
		%\textcolor{red}{FS: Already corrected with the alpha particles included}
	
	%%%%%%%%%%%%%%%%%%%%%	
	%%%%%%
	
For failed CC SN  explosions, we only have two models from different groups, and therefore we do not attempt any interpolation.

	\subsection{Diffuse SN ALP background}
	\label{sec:DSNALPB_theory}
	
	From the SN ALP flux described in the previous section, one can calculate the DSNALPB
	from all past CC SNe in the Universe, as in Paper I (see also~\cite{Beacom:2010kk,Raffelt:2011ft} 
	and in particular Sec.~VI of Ref.~\cite{Caputo:2021rux}
	for a detailed derivation of this equation),
	%...........................................
	\begin{eqnarray}
		\frac{d \phi_a (E_a)}{d E_a}\!\! &=&\!\! \int_0^{\infty} \! \left[ (1+z) \frac{dN_a^{CC}(E_a(1+z))}{dE_a}
		\right]
		\\\nonumber
			%\qquad\qquad\qquad\qquad
		&\times& [R_{CC}(z)] \bigg[ \bigg|c \frac{dt}{dz} \bigg| dz \bigg] .
		\label{eq:diffuse}
	\end{eqnarray}
	%...........................................
	The first term in large brackets is the emission spectrum 
	$dN_a^{CC}/dE_a$, 
	where an ALP received at energy $E_a$ was emitted at a higher energy $E_a(1+z)$; the prefactor of $(1+z)$ on the spectrum accounts for the compression of the energy scale, due to the redshift $z$.
The second term is the supernova rate density $R_{CC}(z)$.
	The third term is the differential distance
	where  $ |{dt}/{dz} |^{-1}= H_0(1+z)[\Omega_\Lambda+\Omega_M(1+z)^3]^{1/2}$ with the cosmological parameters 
	$H_0= 67$ km s$^{-1}$ Mpc$^{-1}$, $\Omega_M=0.3$, $\Omega_\Lambda=0.7$.

	\medskip
	
	{\bf ALP spectrum from past core-collapse SNe.}
	In order to calculate the ALP spectrum of past CC SN events
	$dN_a^{CC}/dE_a$, 
	one has to 
	weight the flux from a given CC SN over the initial mass function (IMF)
	which provides  the number of stars formed per unit of mass as function of the progenitor mass $M$.

	Following Ref.~\cite{Horiuchi:2008jz}, we show the 
	results for three IMFs: a traditional
	Salpeter IMF~\cite{Salpeter:1955it}, an intermediate Kroupa IMF~\cite{Kroupa:2000iv},
	and a shallower Baldry-Glazebrook (BG) IMF~\cite{Baldry:2003xi}.
	The different IMF are characterized by the parameter $\zeta$, defined in the expression below
	%...............
	\begin{equation}
		\phi(M) \propto M^{-\zeta} \,\ .
		\label{eq:imf}
	\end{equation}
	%..................
	For $M\gtrsim 0.5 M_{\odot}$, we find
	$\zeta=2.35$ for the Salpeter IMF,
	$\zeta=2.3$ for the Kroupa case and $\zeta=2.15$ for the BG IMF.
	
	It is expected that the IMF of stars may depend systematically on the environment. In this context, in Ref.~\cite{Gutcke:2019} it was suggested to empirically investigate the effect of metallicity changing the exponent $\zeta$ in a range $[0.34:3.44]$. We find that the effect can produce a factor $\sim 2$ change in the DNSALPB flux.
	
	In our study, we consider masses from 8 up to
	125 $M_{\odot}$. However, due to the steep decline of Eq.~(\ref{eq:imf}), the high-mass end is suppressed 
	and thus of minor relevance for the DSNALPB.
	The IMF-weighted ALP spectrum
	$dN_a^{CC}/dE_a$ of all CC SN events can then be calculated as~\cite{Kresse:2020nto} 
		%.................................
%	\begin{equation}
%		\frac{dN_a^{CC}}{dE_a}=  \frac{
%			\int dM \phi(M) \frac{dN_a}{dE}(M)}
%		{\int_{8 M_{\odot}}^{125 M_{\odot}} dM \phi(M) } \,\ .
%	\end{equation}
\begin{widetext}
		\begin{equation}
		\frac{dN_a^{CC}}{dE_a}= \frac{
		 \int_{\Lambda_{\rm expl-CC}} dM \phi(M) \frac{dN_a}{dE}(M) + \int_{\Lambda_{\rm fail-CC}} dM \phi(M) \frac{dN_a}{dE}(M)}
		{\int_{8 M_{\odot}}^{125 M_{\odot}} dM \phi(M) } \,\ ,
	\end{equation}
	\end{widetext}
	%.......................................
	where $\Lambda_{\rm expl-CC}$ and $\Lambda_{\rm fail-CC}$ represent the domains 
	in the progenitor mass range where one expects to have a successful and a failed CC SN explosion progenitor, respectively.
	In particular, the domain of failed CC SN explosions is defined following~\cite{Moller:2018kpn}:
	%................
	\begin{equation}
		f_{\rm fail-CC} = \frac{\int_{\Lambda_{\rm fail-CC}} dM \phi(M)}{\int_{8 M_{\odot}}^{125 M_{\odot}} dM \phi(M)} \,\ ,
	\end{equation}
	%...................
	and implemented here as an hard cut $M^{\rm min}_{\rm fail-CC}$, which represents the lower mass bound of the failed CC SN explosions domain.
	From here, it also follows that
	$f_{\rm expl-CC}=1- f_{\rm fail-CC}$.

	In order to study the DSNALPB sensitivity to $f_{\rm fail-CC}$
	we consider four different scenarios, as in ~\cite{Moller:2018kpn}. Each scenario is characterized by a different $M^{\rm min}_{\rm fail-CC}$. We consider that all stars with $M > M^{\rm min}_{\rm fail-CC}$ evolve into BH-SNe.
	For progenitor masses lower than $M^{\rm min}_{\rm fail-CC}$ we assume successful explosions with spectrum in Eq.~(\ref{eq:time-int-spec}) following the scaling of the parameters given by Eq.~(\ref{eq:parameters}).
	Instead, we model failed CC SNe explosions as the 40~$M_{\odot}$ model for $ M^{\rm min}_{\rm fail-CC} < M <$ 60~$M_{\odot}$. 
	In the range $[60:125] M_{\odot}$ they are represented by the 
	$70 M_{\odot}$ model.
	The four scenarios are:
	\begin{itemize}
	    %\item[-] \JJC{Show fail/success plot and include the option for the results following from this in the plot. R: it is done later}
		\item[-] $ f_{\rm fail-CC} =10 \%$: $M^{\rm min}_{\rm fail-CC} = 36 M_{\odot}$, see e.g.~\cite{Woosley:2002zz,Pejcha:2014wda}.
		%This corresponds to the assumption that all stars with $M > M^{\rm min}_{\rm fail-CC} = 36 M_{\odot}$ evolve into BH-SNe~\cite{Woosley:2002zz,Pejcha:2014wda}. 
		This is a (as concerns ALPs)  optimistic scenario where the fraction of SNe leading to black hole formation is small. 
		%In this cases all the SN in the range [$36; 60$] are modeled as  the 40~$M_{\odot}$.
		\item[-] $ f_{\rm fail-CC} =20 \%$: $M^{\rm min}_{\rm fail-CC} = 24.0$~$M_{\odot}$, following  Refs.~\cite{Ertl:2015rga,Raithel:2017nlc}.
		%, progenitors with $M > M^{\rm min}_{\rm fail-CC} = 24$~$M_{\odot}$  are assumed to be failed explosions, while for lower masses we assume the scaling of the parameters given by  Eq.~(\ref{eq:parameters}).
		\item[-] $ f_{\rm fail-CC}=30 \%$:   $M^{\rm min}_{\rm fail-CC} = 18.5$~$M_{\odot}$.
		\item[-] $ f_{\rm fail-CC}=40 \%$:   $M^{\rm min}_{\rm fail-CC} = 15.0$~$M_{\odot}$. This is based on the findings of Ref.~\cite{Sukhbold:2015wba,Hidaka:2016zei} and it is still well within the observational constraints~\cite{Adams:2016hit}.
	\end{itemize}

In principle, it has been recently shown that the appearance of exotic phases of hot and dense matter, associated with a sufficiently strong phase transition from nuclear matter to the quark-gluon plasma at high baryon density, can trigger supernova explosions of massive stars in the range $35-50$~$M_{\odot}$. However, from nucleosynthesis studies it results that the contribution of these exotic SNe might be at most 1 \% of the total ones~\cite{Fischer:2020xjl}. Therefore, their contribution to the DSNALPB is negligible and we will neglect it hereafter.	
	
	%.........................	
	%\subsubsection*{Supernova rate $R_{\rm SN}$}
	%.............................
	
	\medskip
	
	{\bf Supernova rate $R_{\rm CC}$}. The intensity and spectrum of the DSNALPB depend on the cosmological rate of core collapse (or, shortly, Supernova Rate, SNR). The SNR, differential in the progenitor mass $M$,  is proportional to the star formation rate (SFR), $R_{\rm SF}(z)$ (defined as the mass that forms stars per unit comoving volume per unit time, at redshift 
	$z$)~\cite{Lunardini:2010ab,Moller:2018kpn}:
	%.......................
	\begin{equation}
		R_{\rm CC}(z,M)= R_{\rm SF}(z) \frac{\int_{8.0 M_{\odot}}^{125 M_{\odot}} dM \phi(M)}{\int_{0.5 M_{\odot}}^{125 M_{\odot}} dM M \phi(M)} \,\ .
	\end{equation}
	%.....................
	The SFR is well described by the functional
	fit~\cite{Yuksel:2008cu}
	%..................
	%.................................................................
	\begin{equation}
	%	\label{sfr} 
		R_{\rm SF}(z) = R_{\rm SF}(0) \biggl[ (1+z)^{\alpha \eta}+\biggl(\frac{1+z}{B}\biggr)^{\beta\eta}+\biggl(\frac{1+z}{D}\biggr)^{\gamma\eta}\biggr]^{1/\eta}\;.
		\label{Eq:RSF}
	\end{equation}
	%.....................................................................
	where $R_{\rm SF}(0)$
	is the normalization
	(in units of $M_{\odot} \,\ \textrm{yr}^{-1}\,\ \textrm{Mpc}^{-3}$)
	,  $B$ and $D$ encode the redshift breaks, the
	transitions are smoothed by the choice $\eta \simeq -10$, and 
	$\alpha$, $\beta$ and $\gamma$ 
	are the logarithmic slopes of the low, intermediate, and high redshift regimes, respectively. The constants $B$ and $D$ are defined as
	%........
	\begin{eqnarray}
	B &=&(1+z_1)^{1-\alpha/\beta} \,\ , \nonumber \\
	D &=& (1+z_1)^{(\beta-\alpha)/\gamma}(1+z_2)^{1-\beta/\gamma} \,\ ,
	\end{eqnarray}
	%............
	where $z_1$ and $z_2$ are the redshift breaks. 
	All the parameters of the model are collected in Tab.~\ref{tab:RSFparameters} based on~\cite{Horiuchi:2008jz}. 
	In the Table the parameters refer to the Salpeter IMF. 
	For the Kroupa and the BG IMFs the normalization factor is reduced by $\simeq 0.94$ and $\simeq 0.76$, respectively, while the overall shape is not greatly affected (see Table 2 of~\cite{Hopkins:2006bw}).
	%...................
		\begin{table}[!t]
		\caption{Model parameters for the SFR, Eq.~(\ref{Eq:RSF}), values taken from~\cite{Horiuchi:2008jz}.
		}
		\begin{center}
			\begin{tabular}{lcccccc}
				\hline
				Analytic fits & 
				\,\ $R_{\rm SF}(0)$ \,\   
				& \,\ $\alpha$ \,\ & \,\ $\beta$  \,\  & \,\ $ \gamma$  \,\  &  \,\ $ z_1$  \,\  &   \,\ $z_2$ \,\ \\
				\hline
				\hline
				Upper  & 0.0213 & 3.6 & -0.1 & -2.5 & 1 & 4\\
				Fiducial & 0.0178 & 3.4 & -0.3 & -3.5 & 1 & 4\\
				Lower & 0.0142 & 3.2 & -0.5 & -4.5 & 1 & 4\\
				\hline
			\end{tabular}
			\label{tab:RSFparameters}
		\end{center}
	\end{table}
	%%%%%%%%%%%%%%%%%%%%%%%%%%%%%%%%
\medskip
	
\begin{figure}[t!]
			\vspace{0.cm}
			\includegraphics[width=0.95\columnwidth]{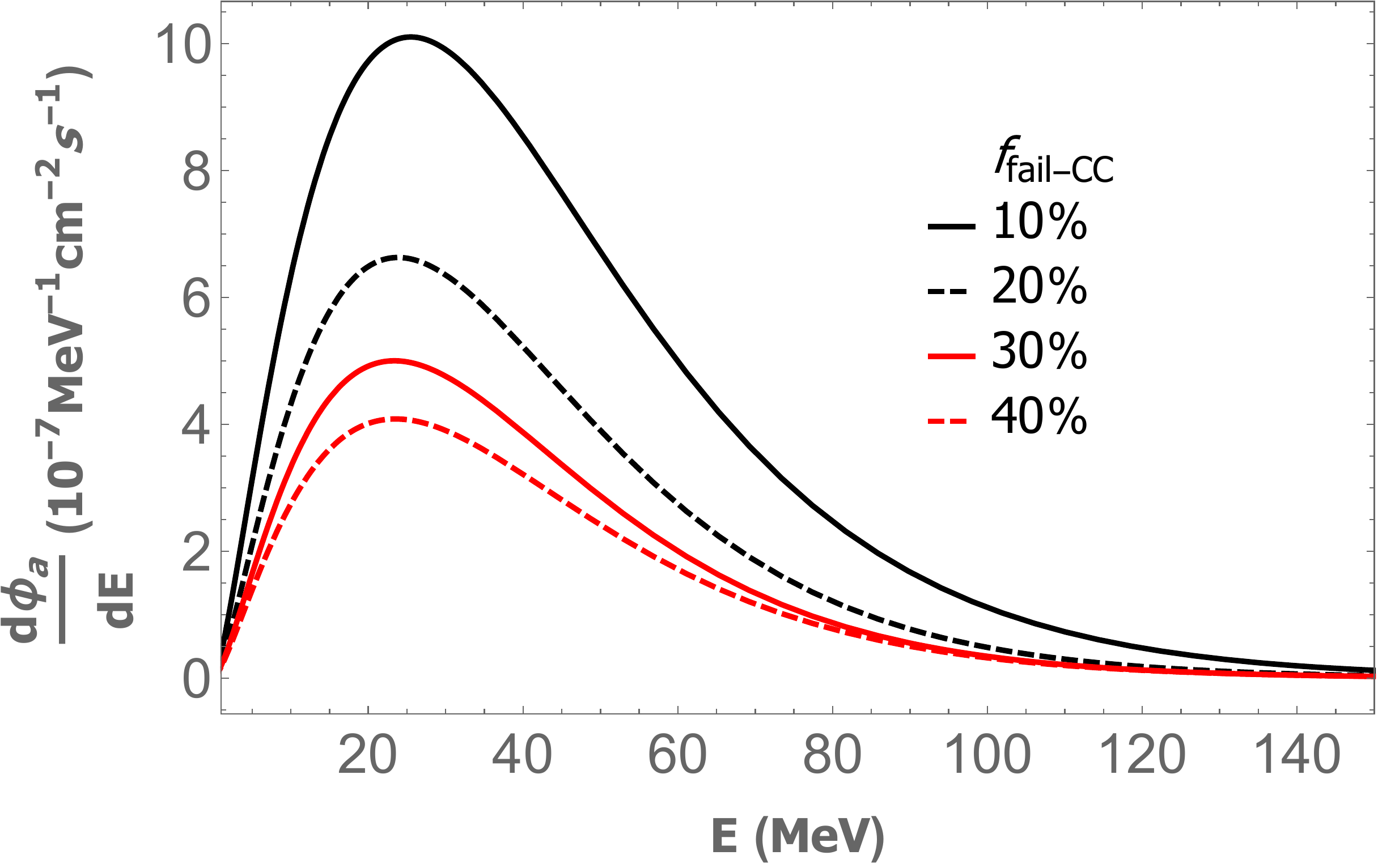}
			\caption{DSNALPB fluxes with  $g_{a\gamma} = 10^{-11}$ GeV$^{-1}$ 
			%\sout{and $m_a \ll 10^{-11}$~eV}
				for different fractions of failed SNe $f_{\rm fail-CC}$, assuming the fiducial model for $R_{\rm SF}$.
			}
			\label{fig:dsalp_ffail}
		\end{figure}

	{\bf DSNALPB flux.}	
	We are now ready to discuss how the different uncertainties in the calculation discussed above impact the DSNALPB flux.
	In Fig.~\ref{fig:dsalp_ffail} we show the DSNALPB fluxes for a  photon coupling $g_{a\gamma} =  10^{-11}$ GeV$^{-1}$
	and $m_a \ll 10^{-11}$~eV for the different fractions of failed SNe $f_{\rm fail-CC}$, assuming the fiducial model
	of Table \ref{tab:RSFparameters} for 
	the $R_{\rm SF}$.
%	We note the obvious 
	As expected, the larger $f_{\rm fail-CC}$ the more suppressed is the flux. The flux uncertainty related to the unknown fraction of failed SNe is 
	a factor $\sim 3$.

		%%%%%%%%%%
		%%%%%%%%%%%%
		
\begin{figure}[t!]
				\vspace{0.0cm}
			\includegraphics[width=0.95\columnwidth]{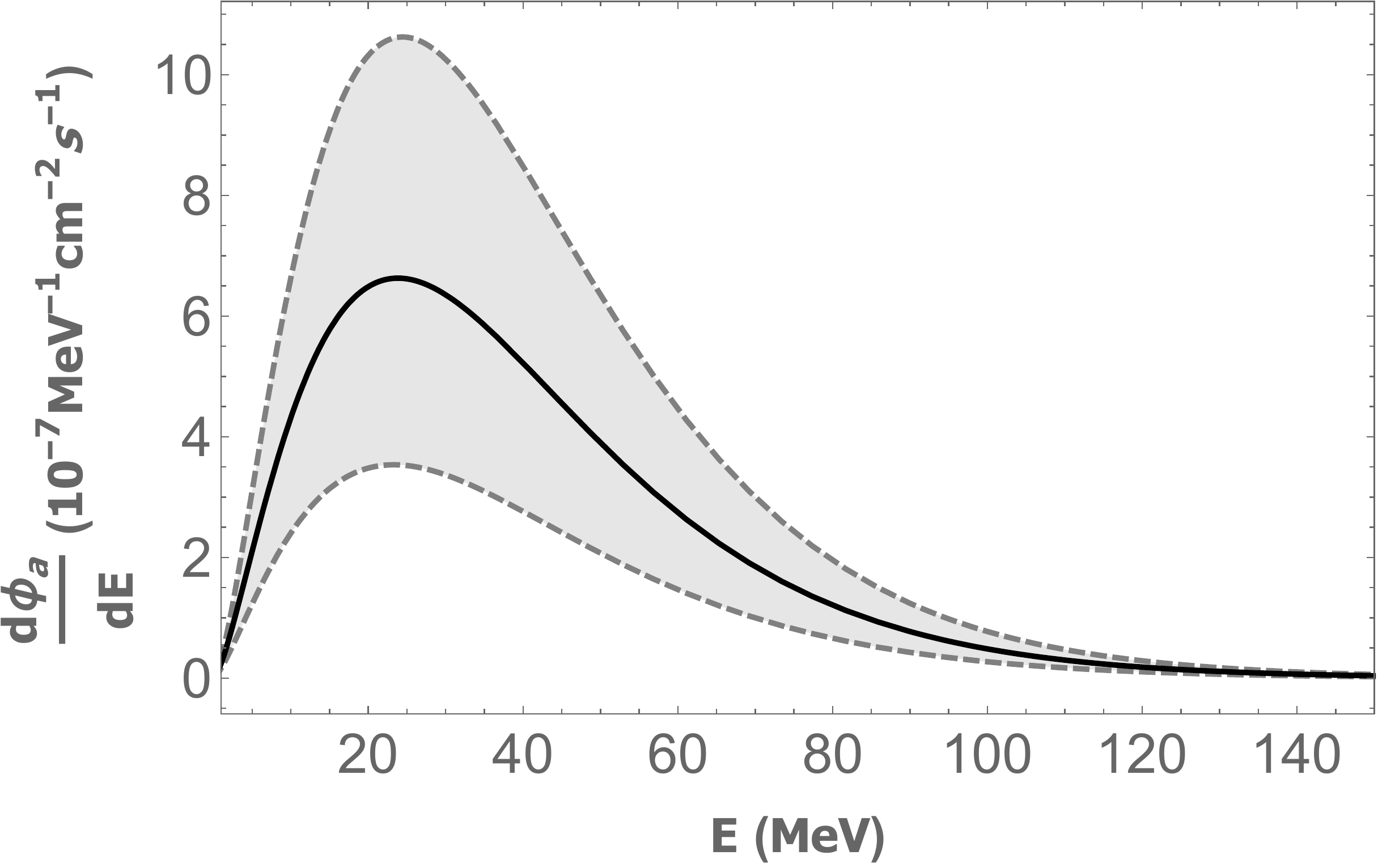}
	\caption{DSNALPB fluxes  for  $g_{a\gamma} = 10^{-11}$ GeV$^{-1}$, 
	%\sout{and $m_a \ll 10^{-11}$~eV} 
	$f_{\rm fail-CC }=20\%$ and the Salpeter IMF.  The   uncertainty range is due to the $R_{\rm SF}$ parameters of  Table~\ref{tab:fitting}.
	The continuous curve refers to the fiducial model, while upper and lower curves refer to upper and lower models, respectively.
			}
			\label{fig:dsalprsf}
		\end{figure}

In Fig.~\ref{fig:dsalprsf}	we show the impact of the changes of parameters in the 
$R_{\rm SF}$ of Table~\ref{tab:fitting}. 
We fix $f_{\rm fail-CC }=10\%$ and Salpeter IMF. The continuous curve refers to the fiducial model for $R_{\rm SF}$, while upper and lower curves refer to upper and lower models, respectively. The uncertainty on $R_{\rm SF}$ leads to a factor $\sim 3$ of variation in the DSNALPB flux.
Instead, the variation associated with a different choice of IMF is subleading.

\begin{figure}[t!]
			\vspace{0.cm}
			\includegraphics[width=0.95\columnwidth]{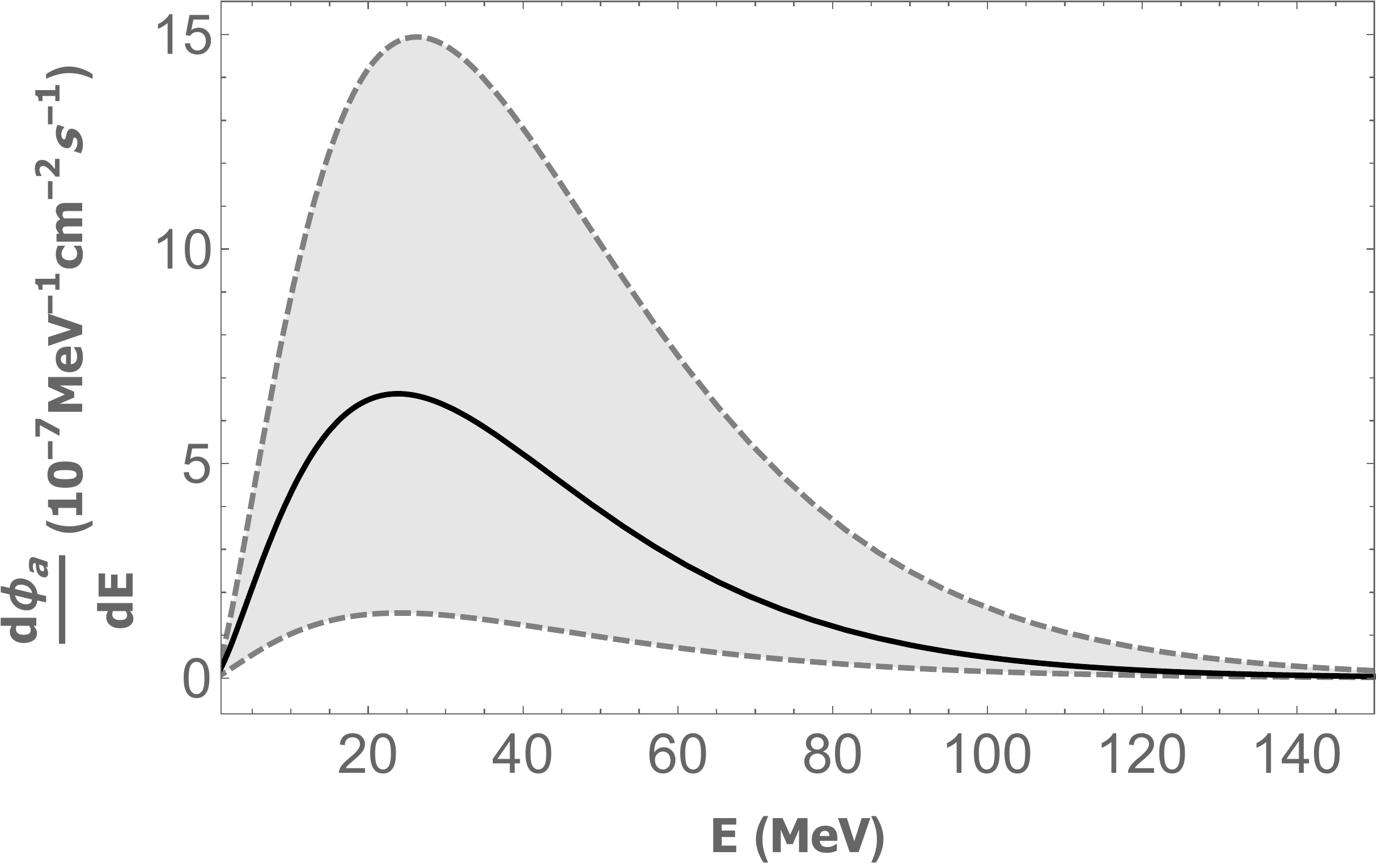}
			\caption{DSNALPB fluxes range of variability (gray band) for  $g_{a\gamma} = 10^{-11}$ GeV$^{-1}$. 
			%\sout{and $m_a \ll 10^{-11}$~eV.} 
			The lower dashed line corresponds to $f_{\rm fail-CC} = 40\%$, BG IMF and lower model parameters for $R_{\rm SF}$ in  Table~\ref{tab:fitting}, while the upper dashed curve corresponds to 
$f_{\rm fail-CC} = 10\%$, Salpeter IMF and upper model parameters for $R_{\rm SF}$. 
The continuous curve refers to $f_{\rm fail-CC} = 20\%$, Salpeter IMF and fiducial model parameters for 
$R_{\rm SF}$.}\label{fig:dsalprange}
		\end{figure}

Finally, we include all the different uncertainties related to the fraction of failed SNe, to the SNR and IMF in order to get a range of variability for the DSNALPB, as shown in the gray band in Fig.~\ref{fig:dsalprange}, where the lower dashed line corresponds to $f_{\rm fail-CC} = 40\%$, BG IMF and lower model parameters for $R_{\rm SF}$ in  Table~\ref{tab:fitting}, while the upper dashed curve corresponds to 
$f_{\rm fail-CC} = 10\%$, Salpeter IMF and upper model parameters for $R_{\rm SF}$. 
For comparison with the continuous curve we show the 
case of $f_{\rm fail-CC} = 20\%$, Salpeter IMF and fiducial model parameters for $R_{\rm SF}$.
We point out that the total range of variability in the DSNALPB flux is a factor
$\sim 8$.

We find that the DSNALPB spectrum can also be represented by the functional form of Eq.~(\ref{eq:time-int-spec}).
	In Table~\ref{tab:fitting} we show the fitting parameters for the four cases with different $f_{\rm fail-CC}$, taking 
	a Salpeter IMF and and a fiducial model
	for the $R_{\rm SF}$ parameters in  Table~\ref{tab:fitting}.
    Typically, in this case the average energy of the spectrum is $E_0 \sim 40$~MeV while its maximum is attained around 25 MeV.
		
			%%%%%%%%%%
		%%%%%%%%%%%%

		%%%%%%%%%%
		%%%%%%%%%%%%

	\begin{table}[!t]
		\caption{
	Fitting parameters for DSNALPB fluxes for $g_{a\gamma} = 10^{-11}$ GeV$^{-1}$ and and $m_a \ll 10^{-11}$~eV for different fractions of failed SNe $f_{\rm fail-CC}$, taking a Salpeter IMF and and a fiducial model for the $R_{\rm SF}$ parameters in  Table~\ref{tab:fitting}. The case
	``max flux'' corresponds to Salpeter  IMF,
	and upper model parameters for $R_{\rm SF}$ in  Table~\ref{tab:fitting}, while the case ``min flux'' corresponds to BG IMF and lower model parameters for $R_{\rm SF}$ in  Table~\ref{tab:fitting}. \label{tab:fitting}
	}
		\begin{center}
			\begin{tabular}{lccc}
				\hline
				$f_{\textrm{fail-CC}}$ & $ $
					$C$ [$\times 10^{-7} \,\  {\rm MeV}^{-1}\textrm{cm}^{-2}\textrm{s}^{-1}$] \,\ 
					& $E_0$ [MeV] &$\beta$ \\
					\hline
					\hline
					10\%  \textrm{max flux}  & $144   \,\   $ & 43.8 &1.50 \\
					10\% & $88.9  \,\     $ & 43.5 &1.41 \\
					20\%  & $62.9  \,\     $ & 39.9 &1.49 \\
					30\%  & $46.5   \,\    $ & 39.3 &1.47 \\
					40\%  & $35.8   \,\    $ & 40.2 &1.41 \\
					40\% \textrm{min flux}  & $15.7  \,\     $ & 42.3 &1.32 \\
					\hline
				\end{tabular}
			\end{center}
		\end{table}

\medskip

{\bf DSNALPB conversions into gamma rays.} ALPs  produced in a SN propagate until they reach the Milky Way, where they  can convert into photons in the Galactic magnetic field (GMF). To calculate the conversion probability we follow the same procedure used in Paper I and  Refs.~\cite{Horns:2012kw}).
%\JJC{Below a new proposal for the summary of previous results, somewhat reducing the overlap with our previous discussion. What do you think?}

%\JJ{To calculate the conversion probability we follow the same procedure that we also used in our previous paper~\cite{Calore:2020tjw} (which in turn was based on~\cite{Horns:2012kw}). To keep the presentation self contained let us nevertheless briefly summarize the main features.}

%\mg{I am not sure we can say "our previous paper", since not all of us worked on that paper. Perhaps, better to say "To calculate the conversion probability we follow the same procedure used in Refs.~\cite{Calore:2020tjw,Horns:2012kw})."} \FC{Agree with MG}

As it is well known (see~\cite{Raffelt:1987im} for the seminal paper discussing this in detail), in a homogeneous magnetic field, ALPs can convert into photons with a polarization parallel to the magnetic field.
For massless ALPs, in vacuum and at sufficiently weak coupling the conversion probability after a length $d$ is, 
\begin{eqnarray}
    P_{a\to \gamma}&=&\left(\frac{g_{a\gamma}B_{T}}{2}\right)^{2}d^2
    \\\nonumber
    &\sim& 0.015 \left(\frac{g_{a\gamma}}{10^{-11}\,{\rm GeV}}\right)^2\left(\frac{B_T}{10^{-6}\,\rm G}\right)\left(\frac{d}{\rm kpc}\right)^2.
\end{eqnarray}
Here, $B_T$ is the magnetic field strength transverse to the propagation direction of the ALP. In the Galaxy we expect fields of the order of \textmu G, see~\cite{Jaffe:2019iuk} for a comprehensive review.
For the chosen value of the coupling we can therefore expect appreciable conversion inside the Galaxy.

However, there are additional effects that have to be taken into account to achieve a realistic description inside the Galaxy.
In the Galaxy neither the strength nor the the direction of the magnetic field is constant. Therefore, one has to integrate the build up of the photon amplitude for both possible polarization directions along the line of propagation through the Galaxy.
We solve the relevant equations numerically. To do so we need the Galactic magnetic field model as an input.
As our baseline model  we take the Jansson-Farrar model~(\cite{Jansson:2012pc}) with the updated parameters given in Tab.~C.2 of~\cite{Planck:2016gdp} (``Jansson12c'' ordered fields)\footnote{We comment that as pointed out in Ref.~\cite{2019ApJ...877...76K} the Jansson and Farrar model  exhibits regions in which the magnetic divergence constraint is violated. Prescriptions have been proposed to mitigate this problem in \cite{2019ApJ...877...76K}. This issue would deserve a dedicated investigation in relation to ALP-photon conversions.}. To quantify the uncertainty due to the magnetic field, we also compare to the the Pshirkov model~\cite{Pshirkov:2011um}. This second model features a larger magnetic field in the Galactic plane and a weaker off-plane component, and, to the best of our knowledge, it is not excluded yet by Faraday rotation data.

The propagation is further complicated by changes in the wavelength of the photon and the ALP. These arise from the mass of the ALP, the plasma mass of the photon arising from the non-vanishing electron density, as well as, indeed, the coupling between the ALP and the photon inside the magnetic field. The ALP mass and the photon coupling are explicit parameters of the ALP model, i.e.~the parameters we want to constrain. The plasma mass is directly related to the electron density which we take as an astrophysical input.
For the electron density,  we use the model described in~\cite{Cordes:2002wz} (for both magnetic field configurations).
In general the effect of the photon and plasma mass on the probability is energy dependent and fully included in our analysis. We note however, that for  $m_{a}\lesssim 10^{-11}\,{\rm eV}$ and $g_{a\gamma}\lesssim 10^{-11}\,{\rm GeV}$ and energies $E\gtrsim 50\,{\rm MeV}$ the mass effects become negligible and the probability is energy independent.

\begin{figure}[t!]
\vspace{0.cm}
\includegraphics[width=0.95\columnwidth]{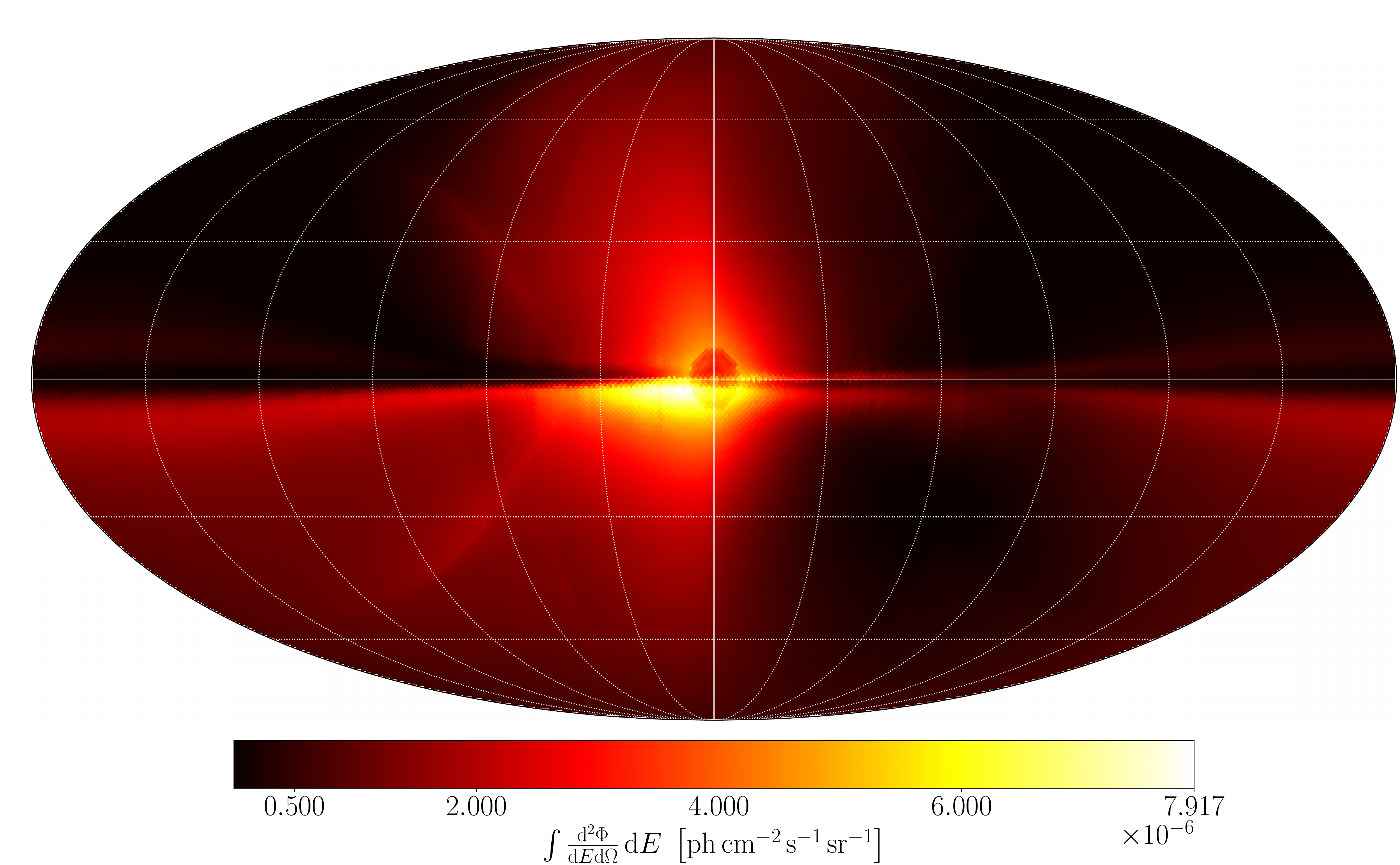}
\caption{All-sky map in Galactic coordinates of the photon flux from the DSNALPB,  $\frac{\mathrm{d}^2\Phi}{\mathrm{d}E\mathrm{d}\Omega}$, integrated from 50 MeV to 200 MeV (corresponding energy range of the low-energy \Fermi-LAT data set used in the following analysis, see Sec.~\ref{sec:data-selection}) with respect to the benchmark scenario defined in Sec.~\ref{sec:results}.
%\sout{For the $a\to \gamma$ conversion probability in the Milky Way, we started from a pure ALPs beam at the outside boundary of the Galaxy, for the Jansson and Farrar magnetic field model derived in~\cite{Jansson:2012pc} and with parameters updated according to \cite{Planck:2016gdp}.}\FC{I moved this in the main text} 
The assumed coupling $g_{a\gamma} = 3.76 \times  10^{-11}$GeV$^{-1}$ (for $m_a \ll 10^{-11}$ eV) represents the $95\%$ CL upper bound derived later in this analysis for the benchmark DSNALPB scenario (cf.~Sec.~\ref{sec:results}). %{\color{cyan}{FS: Maybe the lower label needs to be bigger.}}
}
\label{fig:gal}
\end{figure}

In Fig.~\ref{fig:gal} we show an example of the all-sky DSNALP gamma-ray flux, resulting from the numerical implementation of the procedure outlined above.
For the $a\to \gamma$ conversion probability in the Milky Way, we started from a pure ALPs beam at the outside boundary of the Galaxy, for the Jansson and Farrar magnetic field model derived in~\cite{Jansson:2012pc} and with parameters updated according to \cite{Planck:2016gdp}. 
Besides giving an  idea of the magnitude of fluxes at play from ALPs, this map represents the spatial distribution of the signal\footnote{Due to the energy independence of the conversion in the energy range of Fermi LAT and for the chosen ALP mass parameters the morphology is actually the same as in Fig.~3 of Paper I, where the conversion probability and not the signal was plotted. There, however, only the average was used to set the limits.} as it is used, for the first time in this work, as input for the \Fermi-LAT analysis.

\section{Fermi-LAT Analysis framework}
\label{sec:data-analysis}

\subsection{Data selection}
\label{sec:data-selection}

We use 12 years of \Fermi-LAT Pass8 data. The signal is peaked at about 25~MeV. Therefore, we use two separate
data sets with different selection criteria to specifically improve
the analysis of LAT data below 200 MeV. The applied criteria are summarized
in Tab.~\ref{tab:LAT-data-selection}. 

\begin{table*}[t]
\begin{centering}
\begin{tabular}{|l||c|c|}
\hline 
{Data Set} & {$E<200$ MeV} & {$E\geq200$ MeV} \tabularnewline
\hline 
\hline 
\hline
{Reconstruction algorithm} & \multicolumn{2}{c|}{{Pass 8}}\tabularnewline
\hline 
{Event class} & \multicolumn{2}{c|}{{ULTRACLEANVETO}}\tabularnewline
\hline 
{Event type} & {PSF3} & {FRONT+BACK}\tabularnewline
\hline 
{Energy range}  & {50 MeV - 200 MeV} & {200 MeV - 500 GeV}\tabularnewline
\hline 
{Time interval} & \multicolumn{2}{c|}{{\makecell{12 years (4th August 2008  \\ - 3rd September 2020)}}}\tabularnewline
\hline 
{ROI} & \multicolumn{2}{c|}{{all sky}}\tabularnewline
\hline 
{Zenith angle (applied to gtltcube)} & \multicolumn{2}{c|}{{$<90^{\circ}$}}\tabularnewline
\hline 
{Time cuts filter} & \multicolumn{2}{c|}{{\makecell{DATA\_QUAL==1 \&\& \\ LAT\_CONFIG==1}}}\tabularnewline
\hline 
{HEALPix resolution} & \multicolumn{2}{c|}{{$N_{\mathrm{side}}=64$}}\tabularnewline
\hline 
{energy binning} & \multicolumn{2}{c|}{{30 logarithmically-spaced bins}}\tabularnewline
\hline 
\end{tabular}
\par\end{centering}
\caption{Data selection and preparation specifications \label{tab:LAT-data-selection}.}
\end{table*}

While the $E<200$ MeV data is the main driver of the constraint, let us nevertheless start by describing our procedure for the more standard $E\geq200$ MeV data set. This gives the picture of the main ingredients in our analysis. We will later comment on the adaptations for the $E<200$ MeV region.

The data set of events $E\geq200$ MeV includes both front- and back-converted
events to increase the statistical sample, whereas the data set with
$E<200$ MeV is restricted to photons of the PSF3 event type. This
decision has been made to benefit from the slightly better angular
reconstruction efficiency of this event type compared to the generally
poor angular resolution of the LAT at the lower end of its sensitivity
range\footnote{The description of the LAT's performance figures can be found at \url{https://www.slac.stanford.edu/exp/glast/groups/canda/lat_Performance.htm}}. For both data sets the \texttt{ULTRACLEANVETO} event class
has been chosen as it minimizes the contamination by misclassified
cosmic-ray events, which is essential for studies of large-scale diffuse
sources like the extragalactic ALP flux from SNe. The choice of this
event class requires us to select the \Fermi-LAT Instrument Response
Functions (IRFs) \texttt{P8R3\_ULTRACLEANVETO\_V3} with which we
will convolve the physical gamma-ray emission models to generate from
them the expected number of photon events. The LAT data as well as
the model data to be generated are stored as all-sky maps and binned
according to the HEALPix pixelization scheme \cite{2005ApJ...622..759G} with $N_{\mathrm{side}}=64$.
The mean distance between the centers of two such HEALPix pixels amounts
to about $0.9^{\circ}$. All data manipulations involving either the
LAT data or the application of LAT IRFs is done via the Fermi Science
Tools\footnote{\url{https://github.com/fermi-lat/Fermitools-conda}}
(version 2.0.8).

\subsection{Methodology}
\label{sec:methodology}

The ALP-induced gamma-ray flux manifests itself as a large-scale contribution
to the overall gamma-ray sky at low Galactic latitudes around the
Galactic disc as well as at high Galactic latitudes. To do justice
to this fact, we develop a template-based analysis that utilizes all-sky
maps of the expected photon counts for various background components
and the ALP signal template. The selection of astrophysical gamma-ray
emission backgrounds comprises Galactic and extragalactic contributions
that are commonly considered in studies of the LAT data. To give a
rough outline of the analysis strategy, we first single out the region
of the sky that yields the best agreement between a model built from
the astrophysical emission components. In a second step, this ROI is used to constrain the strength of the ALP-induced gamma-ray
flux.

\medskip
\textbf{Astrophysical background model selection.} The model for the
gamma-ray sky is created from a selection of the ``guaranteed''
emission components on which we comment in the following. We process
these models with the Fermi Science Tools and its dedicated routines;
in particular, the routine \textit{gtmodel} to derive photon count
templates, i.e.~templates convolved with the LAT's PSF\footnote{\href{https://www.slac.stanford.edu/exp/glast/groups/canda/lat_Performance.htm}{Fermi-LAT key performance figures}}
and multiplied by the exposure depending on the data set (see Tab.~\ref{tab:LAT-data-selection})
to obtain the ``infinite statistics'' or \emph{Asimov} dataset \cite{Cowan:2010js}.
We incorporate in the analysis:
\begin{itemize}
\item Interstellar emission (IE) -- the combined gamma-ray flux due to
high-energy charged cosmic rays interacting with gas, photon radiation
fields and dust in the Milky Way -- which is represented by five
distinct models to examine the robustness of the analysis with respect
to variations of this particular component. From the wide range of
different attempts to quantify the intensity, spatial and spectral
structure of the Galactic IE, we choose as the
benchmark in our analysis one particular model instance that has been
created to examine the systematic uncertainty inherent to the ``1st
Fermi LAT Supernova Remnant Catalog'' \cite{Acero:2015prw}.\footnote{The model files have been made public by the \Fermi-LAT collaboration: \url{https://fermi.gsfc.nasa.gov/ssc/data/access/lat/1st_SNR_catalog/}. We note that these files have been initially generated to be compatible with Pass7 LAT data. However, they may be manually converted to comply with the Pass8 standard by using the same factor that distinguishes the official \Fermi-LAT diffuse background models \href{https://fermi.gsfc.nasa.gov/ssc/data/analysis/software/aux/gll_iem_v05.fits}{\texttt{gll\_iem\_v05}} and \href{https://fermi.gsfc.nasa.gov/ssc/data/analysis/software/aux/gll_iem_v06.fits}{\texttt{gll\_iem\_v06}}.} In what
follows, we will refer to this model by ``Lorimer I''. While the
documentation of the exact details of this model can be found in the
referenced publication~\cite{Acero:2015prw}, we stress here the basic assumptions underlying
its construction: The sources of primary cosmic rays are assumed to
follow the distribution of pulsars in the Milky Way as reported in
\cite{Lorimer:2006qs}. The typical height of the cosmic-ray propagation halo
is set to $z=10$~kpc, while the spin temperature of the interstellar
medium is taken to be $T=150$ K. These model parameters and assumptions
are not largely different from similar models that have in the past
and recently been applied to study the characteristics of the gamma-ray
emission in the Galactic center region \cite{TheFermi-LAT:2017vmf,DiMauro:2021raz}.
Another advantage of this model is its decomposition into an inverse
Compton map and gas maps (notably atomic \ce{H} as well as \ce{CO}
as a proxy for the distribution of \ce{H2}), which are themselves
split into Galactocentric annuli of various extension (0-4 kpc: ``ring 1'', 4-8
kpc: ``ring 2'', 8-10 kpc: ``ring 3'' and 10-30 kpc: ``ring 4''). This subdivision into annuli allows
us to perform an optimization of the individual model components via
an all-sky baseline fit which we describe later. \linebreak{}
We complement this benchmark choice with four additional interstellar
emission models (IEMs): ``Lorimer II'' -- another model instance from
\cite{Acero:2015prw} with the only difference from Lorimer I being
an extreme choice for the spin temperature which is taken to be $T=1\cdot10^{5}$
K as well as the ``Foreground Models'' A, B and C from the in-depth
Fermi-LAT study of the diffuse extragalactic gamma-ray background
\cite{Ackermann:2014usa}.\footnote{The relevant model files can be retrieved from the \Fermi-LAT collaboration's
public data archive: \url{https://www-glast.stanford.edu/pub_data/845/}} The IEMs of the latter publication possess
the advantageous feature of having been created with the idea in mind
that they will eventually be used to study high-latitude LAT data;
a task that we are likewise aiming at.
\item Isotropic diffuse background (IGRB) -- The spatial morphology of
this component follows the exposure of the LAT while its spectrum
is determined in connection with a particular IEM. For our analysis, we adopt the IGRB component shipped with
the Fermi Science Tools\footnote{The relevant spectrum files are also provided at \url{https://fermi.gsfc.nasa.gov/ssc/data/access/lat/BackgroundModels.html}}
and respecting the choice of event class and type in the context of
the two data sets in Tab.~\ref{tab:LAT-data-selection}. Note that
-- due to reasons that will become clear later while describing the
analysis routine -- the adopted spectrum of the IGRB does not play
a crucial role in our study.
\item Detected point-like and extended gamma-ray sources (PS) -- A \Fermi-LAT
analysis of 10 years of data has revealed more than 5700 individual
gamma-ray sources inside and outside of the Milky Way \cite{Fermi-LAT:2019yla,Ballet:2020hze}.
We include this latest iteration of a high-energy gamma-ray source
catalog, the 4FGL-DR2, in our analysis. Depending on the analyzed
data set, the treatment and handling of these detected sources may
differ and the explicit description of our approach follows later in the text.
\item Fermi Bubbles (FBs) -- As a large-scale diffuse component that extends
to high-latitudes in the northern and southern hemisphere of the projected
gamma-ray sky, we incorporate the FBs as a template according to their
spatial characterization provided in \cite{TheFermi-LAT:2017vmf}.
We adopt as their fiducial spectrum a log-parabola $\frac{\mathrm{d}N}{\mathrm{d}E}=F_{0}\left(\frac{E}{E_{0}}\right)^{-\alpha-\beta\ln\!{\left(E/E_{0}\right)}}$
with parameters $F_{0}=5\times10^{-10}\;\mathrm{ph}\,\mathrm{cm}^{-2}\,\mathrm{s}^{-1}\,\mathrm{MeV}^{-1}$,
$\alpha$ = 1.6, $\beta$ = 0.09 and $E_{0}$ = 1 GeV taken from \cite{Herold:2019pei}.
\item LoopI -- Another large-scale diffuse emission component, which is
most prominently present in the northern hemisphere above the Galactic
disc. We adopt the geometrical spatial structure (and spectral) as
considered in the 1st Fermi-LAT SNR catalog analysis \cite{Acero:2015prw}
that is based on a study in \cite{Wolleben:2007pq}. 
\item Gamma-ray emission from the Sun and the Moon (SUN) -- Both the Sun
and the Moon can contribute a sizeable gamma-ray background when they
pass through the ROI of a particular analysis. Since
we are aiming to conduct an all-sky study, their emission must be
taken into consideration. The Fermi Science Tools offer routines\footnote{An explanation is provided under \url{https://fermi.gsfc.nasa.gov/ssc/data/analysis/scitools/solar_template.html}}
to calculate a LAT data-based Sun and Moon gamma-ray template via
the techniques presented in \cite{2013ICRC...33.3106J}. 
\end{itemize}
\textbf{Statistical inference procedure.} The grand scheme of this
analysis is an all-sky template-based fit. To this end, we construct
a fitting routine that utilizes the Poisson likelihood function subdivided
into energy bins $i$ and spatial pixels $p$\begin{equation} 
\mathcal{L\!}\left(\left.\bm{\mu}\right|\bm{n}\right)=\prod_{i,p}  \frac{\mu_{ip}^{n_{ip}}}{\left(n_{ip}\right)!}e^{-\mu_{ip}} \end{equation}for binned model data $\bm{\mu}$ and experimental data $\bm{n}$.
The model data are a linear combination of the templates $\bm{X}$
introduced above
\begin{equation}
\bm{\mu}=G_a \bm{X}^{\mathrm{ALP}}+\sum_{X}\sum_{i}A_{i}^{X}X_{i}\label{eq:model_eq}
\end{equation}
where $X\in\{\mathrm{IE},\mathrm{IGRB},\mathrm{PS},\mathrm{FB},\mathrm{LoopI},\mathrm{SUN}\}$.
This construction introduces two kinds of normalization parameters. The first are  a set of normalization parameters,
$A_{i}^{X}$, for each energy bin of each astrophysical background component.
These parameters can be varied independently of each other
during a fitting step. The advantage of such an approach is that spectral
imperfections of the original astrophysical emission models are less
impactful as they are re-adjusted in a fit. Thus, a greater emphasis
is given to the spatial morphology of the background components. This
technique has been successfully applied in previous studies, e.g.~\cite{TheFermi-LAT:2017vmf, Macias:2019omb}.
Second, the signal component, i.e.~the ALP-induced gamma-ray
flux, is modelled with a single, global normalization parameter $G_a$
since we aim to exploit both the spatial and spectral shape of this
component. To re-iterate the discussion of the ALP signal in Sec.~\ref{sec:DSNALPB_theory},
its spectral shape is dictated by the physics of core-collapse SNe
while the spatial morphology is a direct consequence of the shape
of the GMF of the Milky Way. Note that while the
importance of the spectral shape of each background component is reduced,
a similar statement about the ALP signal's spectrum is not correct.
Therefore, we need to include energy dispersion\footnote{The Pass 8 reconstruction standard has revealed that energy dispersion
effects are a crucial ingredient to realistically simulate LAT observations.
More information on this subject are provided at this website: \url{https://fermi.gsfc.nasa.gov/ssc/data/analysis/documentation/Pass8\_edisp\_usage.html} } during the generation of the signal template with the Fermi Science
Tools. The impact of energy dispersion is growing with decreasing
photon energy and highly recommended at energies below 100 MeV. Therefore,
we use \texttt{edisp\_bins=-2} (two additional energy bins are added
below and above the nominal energy range of the data set to compute
spectral distortions due to energy dispersion effects) for the data
set of $E<200$ MeV and \texttt{edisp\_bins=-1} for the data set of
$E\geq200$ MeV with \texttt{apply\_edisp=true} in the spectrum part
of the input to the Fermi Science Tools.

We infer the best-fit parameters of the model with respect to one
of the LAT data sets via the maximum likelihood method for which we
invoke the weighted logarithmic Poisson likelihood \cite{Fermi-LAT:2019yla} 

\begin{equation}
\ln\mathcal{L}_{w}\left(\left.\bm{\mu}\right|\bm{n}\right)=\sum_{i,p}w_{ip}\left(n_{ip}\ln\mu_{ip}-\mu_{ip}\right).\label{eq:weighted_likelihood}
\end{equation}
This weighted log-likelihood function has been introduced by the \Fermi-LAT
collaboration in connection with the generation of the 4FGL catalog
as to incorporate the impact of systematic uncertainties on the analysis
results. The basic idea is to assign to each pixel (per energy bin)
a weight -- a quantity that is essentially obtained via integration
in space and energy of the provided model or LAT data -- in order
to suppress the statistical impact of certain parts of the target
region where the emission is dominated by systematic uncertainties.
An exhaustive discussion of the calculation and properties of these
weights can be found in Appendix B of \cite{Fermi-LAT:2019yla}\footnote{Another source that explains this weighted likelihood approach is
found at: \url{https://fermi.gsfc.nasa.gov/ssc/data/analysis/scitools/weighted_like.pdf}}. The numerical routines (\texttt{gteffbkg}, \texttt{gtalphabkg},
\texttt{gtwtsmap}) to compute the weights for a particular setup are
part of the Fermi Science Tools.

As concerns this analysis, we choose to incorporate ``data-driven''
weights in our analysis pipeline. These weights are directly computed
from the selected LAT data. Hence, they yield a means to penalize
pixels that suffer from systematic effects like misclassified charged
cosmic-ray events, PSF calibration, IE spectral modeling uncertainties
in bright regions of the sky or sky parts hosting particularly bright
point-like sources that overshine their surroundings. We fix the level
of the assumed systematic uncertainties to 3\% (for all energy bins),
which is the fiducial value utilized and tested for the creation of
the 4FGL source catalog \cite{Fermi-LAT:2019yla}. The likelihood
maximization step is performed with the \textsc{iminuit} python package
\cite{iminuit} and the migrad minimization algorithm it provides. 

\newpage

To discriminate between different hypotheses -- quantifying a possible
preference for the model in Eq.~\ref{eq:model_eq} with or without
an ALP emission component -- we employ the log-likelihood ratio test
statistic 
\begin{widetext}
\begin{equation} 
\label{eq:TS_stat_reach} 
\textrm{TS}\!\left(G_a\right)= \begin{cases} -2\min_{\{A_i^X\}}\left(\ln\!\left[\frac{\mathcal{L}_w\!\left(\left.\bm{\mu}(G_a,A_i^X) \right|\bm{n}\right)}{\mathcal{L}_w\!\left(\left.\bm{\hat{\mu}}\right|\bm{n}\right)}\right]\right)\, & G_a \geq \hat G_a\\ 0 & G_a < \hat G_a 
\end{cases} 
\end{equation}
\end{widetext}
by adopting the construction discussed in \cite{Cowan:2010js}. In our case at hand, the astrophysical background normalization parameters are treated as nuisance
parameters and $\hat{\cdot}$ denotes the best-fit values of signal
and background normalization parameters. In the case of no significant ALP signal,
this test statistic allows us to set upper limits on the ALP normalization
parameter. As Eq.~(\ref{eq:TS_stat_reach}) only depends on one
parameter and values of $G_a$ smaller than the best-fit value are discarded, its distribution follows a half-$\chi^{2}$-distribution with
one degree of freedom (see Sec.~3.6 of \cite{Cowan:2010js}). Consequently (still following the calculations in the mentioned reference), an $95\%$ CL upper limit on $G_a$ can be set where
the test statistic attains a value of 2.71.

\medskip
\textbf{Fitting procedure.} To derive an upper limit on the strength
of the ALP-induced gamma-ray flux, we have to face and solve two main
challenges:
\begin{enumerate}
\item What is the part of the sky that yields the best agreement between
a model consisting of the six emission components introduced in the
previous section and the measured LAT data? Only such an ROI can be exploited in order to constrain the ALP signal strength
in a statistically sound approach. The manner in which this optimization process is performed was inspired by the approach presented in \cite{Zechlin:2017uzo}, where the authors aim to constrain weakly interacting massive particles via a gamma-ray signal from the Milky Way's outer dark matter halo.
\item How do we have to adapt our fitting procedure to the particular case
of the two data sets above and below 200 MeV? The main concern of
the data set below 200 MeV is the large PSF size of the instrument,
which heavily impacts the manner to incorporate the population of
detected gamma-ray sources from the 4FGL catalog.
\end{enumerate}
The subsequent paragraphs are presenting the reasoning that applies
to the LAT data set above 200 MeV. After this general outline of our
approach, we comment on the parts that need to be altered when handling
the data set below 200 MeV. 

\begin{figure*}[t]
\begin{centering}
\includegraphics[width=0.9\linewidth]{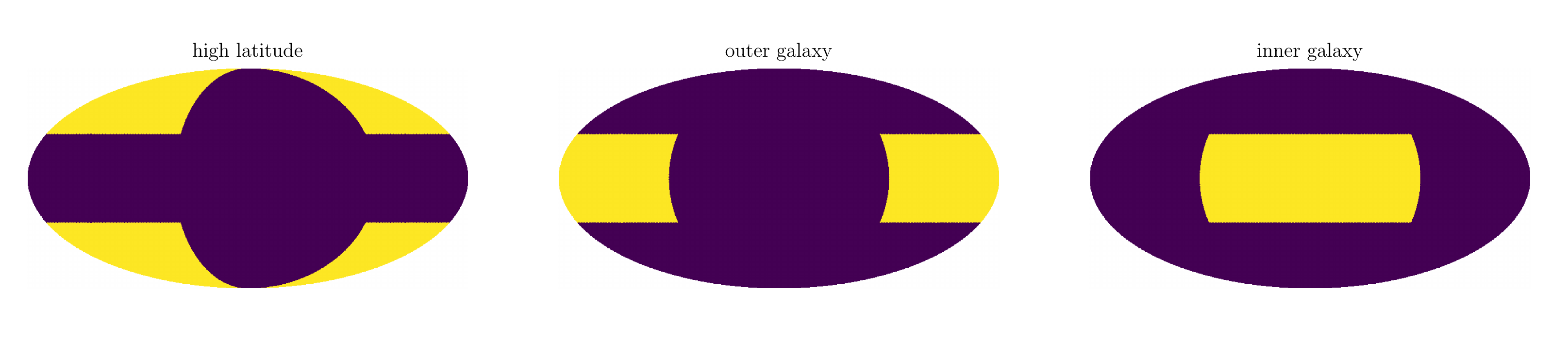}
\par\end{centering}
\caption{The ROIs (yellow) used for the iterative all-sky fitting
pipeline, in Mollweide projection; see text for the definitions of
the regions. \label{fig:The-regions-of}}
\end{figure*}

To answer the first point raised, we adopt and adapt the iterative
all-sky fitting strategy that has been proposed and applied by the
Fermi-LAT collaboration to derive the current iteration of their Galactic
diffuse background model\footnote{The model file can be downloaded from this website: \url{https://fermi.gsfc.nasa.gov/ssc/data/access/lat/BackgroundModels.html}}.
In the companion publication\footnote{\url{https://fermi.gsfc.nasa.gov/ssc/data/analysis/software/aux/4fgl/Galactic_Diffuse_Emission_Model_for_the_4FGL_Catalog_Analysis.pdf}}
that describes the details of the collaboration's analysis, an outline
of the general procedure is given in Sec.~4: The main idea is to
perform a maximum likelihood fit utilising Eq.~\ref{eq:weighted_likelihood}
(and fixed $G_a=0$) by selecting characteristic sky
regions where only a few components would dominate while the sub-dominant
components remain fixed to initial normalization values or the results
of previous iteration rounds. In what follows, we list the definitions
of the different sky regions that we consider in our work and 
those templates -- with respect to our benchmark IEM ``Lorimer I''
-- that are left free therein (masks corresponding to the chosen
regions are shown in Fig.~\ref{fig:The-regions-of}):
\begin{enumerate}
\item \textit{High-latitude}: $|b|>30^{\circ}$ and without the ``patch''-region,
which we define as $-105^{\circ}\leq\ell\leq60^{\circ}$. The patch
region is the part of the sky where LoopI and the FBs are
brightest. Here, we leave free the following templates: HI ring 3,
IC, 4FGL, IGRB and Sun. 
\item \textit{Outer galaxy}: $|b|\leq30^{\circ}$, $|\ell|>90^{\circ}$.
This concerns the following templates: 4FGL, HI ring 4, CO ring 4 and
IC.
\item \textit{Inner galaxy}: $|b|\leq30^{\circ}$, $|\ell|\leq90^{\circ}$.
This concerns the following templates: 4FGL, HI ring 1, HI ring 2,
CO ring 1, CO ring 2, CO ring 3 and IC.
\item \textit{Patch region/all-sky}: To adjust the normalization parameters
of the LoopI and FB templates, we fit them on the full sky while all
other templates are fixed.
\end{enumerate}

After iterating this procedure 100 times, we have obtained a \emph{baseline fit} to the LAT data with which we perform the tests of statistical
robustness of certain ROIs in the following. Moreover,
this routine provides us with a data-optimized IEM that we create
by summing the gas and IC components with their respective best-fit
normalization factors. To avoid fitting all gas rings every time,
we use this optimized IEM as a single template in what follows. Note
that only the IEMs ``Lorimer I'' and ``Lorimer II'' enable
a fit with split gas rings whereas foreground models A, B and C are
treated differently. To conduct the baseline fit in their case, we
split the single IE template into three independent parts coinciding
with the definitions of the sky patches of the iterative fit. The
same reasoning is also applied to the IC template for all five IEMs.

\emph{Region of interest optimization.} Consequently, we systematically search for an ROI that
provides statistically reliable upper limits on the ALP signal's strength.
To this end, we exclusively resort to the southern hemisphere as to
avoid possible contamination by the gamma-ray emission of the rather
poorly constrained Loop I - structure. In addition -- to reduce the human bias on the optimization process of the ROI -- we exchange the physical gamma-ray spectrum of the diffuse ALP background with a simple power law of spectral index -2.\footnote{We ensure that the integrated spatial part of the map is equal to 1 (normalized), resulting in a power law flux normalization $N_0 = 5.8\times10^{-11}$ cm$^{-2}$s$^{-1}$MeV$^{-1}$sr$^{-1}$ at a reference energy of $E_0 = 100$ MeV.}
We fix its reference flux normalization $A^{\mathrm{ALP}}$ so that the resulting flux is one order of magnitude lower than the DSNALPB at a reference energy of 100 MeV and ALP-photon coupling of $g_{a\gamma} = 5.3\times10^{-11}$ GeV$^{-1}$ corresponding to the limit derived in Paper I. Consequently, the maximal photon counts per pixel are of order unity at this reference energy.
%to match the expected order of magnitude of the physical flux.
%\FC{can we say it matches the integrated alps flux from our benchmark model?}
By invoking Eq.~\ref{eq:TS_stat_reach} (replacing $G_a \rightarrow A^{\mathrm{ALP}}$)
and including the ALP template with a non-zero normalization, we derive
the associated TS-distribution in a particular region of the sky,
which we systematically shrink from $\ell\in\left[-180^{\circ},180^{\circ}\right]$
to $\ell\in\left[-90,90^{\circ}\right]$ with $b\in\left[-90^{\circ},-30^{\circ}\right]$.
The cut in Galactic latitude is applied to reduce the impact of the
IE along the Galactic disc. For all tested sky regions, we compare
the resulting TS-distributions for input data $\bm{n}$ that are either
a particular LAT data set or the baseline fit data with respect to
the IEM Lorimer I. The latter data set has to advantage of allowing
us to draw Poisson realizations that eventually show the scatter of
the expected upper limits on $A^{\mathrm{ALP}}$. 
This optimization procedure leads us to the choice of the ROI
for the analysis, presented in Sec.~\ref{sec:results}.
We stress that $A_{\rm ALP}$ is an auxiliary parameter
whose baseline value is connected to the ALPs expected gamma-ray brightness, and used
to tune the analysis pipeline.

\emph{Treatment of detected sources in the 4FGL catalog.} Besides
the $\left(\ell,b\right)$-mask to inspect the admissibility of a
particular ROI in the southern hemisphere, we are also
masking the positions of all detected gamma-ray sources that are listed
in the 4FGL catalog. Each source is masked in a circular region
centered on their nominal position in 4FGL with a radius that corresponds
to the 95\% containment radius of the LAT's PSF at a given energy.
The source mask radius is extended by the reported extension of a
source when applicable. However, this reasoning would lead to masking
the entire sky at energies $E\lesssim500$ MeV. Hence, we only use
the 68\% containment radius for the respective energy bins. We have
checked that increasing the mask radii at these energies does not
impact the final results.

\bigskip

%\JJC{To what degree do we actually make use of the $>200$~MeV Data?} \CE{They are, of course, not the dominant driver of the upper limits. However, this data set does add constraining power for some combinations of Galactic magnetic field and Galactic diffuse model. There is also the energy dispersion of the LAT, which might mitigate some of the "low energy" photons to higher energies thereby broadening the spectrum.}
\emph{Adjustments for the data set $E\leq200$ MeV.} While the overall
rationale of the fitting procedure remains the same, there are a number
of necessary changes to be made in order to optimize the analysis
pipeline at the lowest energies accessible to the Fermi LAT. The LAT's
PSF size rapidly deteriorates at these energies to values larger than
one degree. On the one hand, while bright gamma-ray sources can still
be identified as individual sources, the vast majority of sources
listed in 4FGL will create a sea of photons that rather seems to be
of a diffuse origin and, thus, increasing the degeneracy with genuinely
diffuse signals like the ALP-induced gamma-ray flux. On the other
hand, the ALP signal's spectrum attains its maximal values around
50 to 100 MeV so that this energy range is expected to yield the strongest
constraints.

To account for these obstacles, we first modify the baseline fit routine:
Instead of using a single 4FGL template that encompasses all detected
sources, we split the full template into eight individual templates
defined by the number of expected photons per source $N_{\gamma}$ within the
energy range of the LAT data set, i.e.~$E\in\left[50,200\right]$MeV.
The estimate of $N_{\gamma}$ per source follows from the best-fit
spectrum as reported in the catalog and the LAT exposure. The defining
lower and upper boundaries of each template are:
\begin{itemize}
\item $N_{\gamma}<1$, 
\item $1\leq N_{\gamma}<10$, 
\item $10\leq N_{\gamma}<50$, 
\item $50\leq N_{\gamma}<100$, 
\item $100\leq N_{\gamma}<200$, 
\item $200\leq N_{\gamma}<500$, 
\item $N_{\gamma}\geq500$ without the ten brightest sources,
\item extended sources,
\item each of the ten brightest sources is fit individually.
\end{itemize}
Since the brightest sources in the gamma-ray sky may substantially
impact the quality of the fit, we single out the ten brightest sources
below 200 MeV and fit them individually with the rest of the aforementioned
4FGL templates -- leaving their normalization free only in those
regions of the iterative fit where they are present. After the baseline
fit has converged these ten sources are added to the template with
$N_{\gamma}\geq500$. The resulting all-sky baseline fit and IEM data
is henceforth utilized in the same way as it was done in the case of the
first LAT data set. 

A second adjustment concerns the systematic search for a suitable
ROI: This data set consists of 30 energy bins -- mainly
to guarantee a sufficient sampling of the LAT IRFs and energy dispersion.
While the baseline fit has been conducted with the full number of
energy bins, we rebin this data set to larger macro bins in all later
stages of the analysis. The number of macro bins is a hyperparameter
that needs to be optimized, too. Moreover, only a small fraction of
the detected sources can be masked. The idea is to define a threshold
for each energy bin in terms of number of expected emitted photons
$N_{\mathrm{thr},i}$. If a source exceeds this number, it has to
be masked with a circular mask at 95\% containment radius of the LAT's
PSF\footnote{The PSF size in each macro bin is evaluated at the lowest energy among
the micro bins that are contained in it.}. This will have an impact on the compatibility of LAT data and the
baseline fit data. Hence, we scan over different high-latitude ROI
masks as well as different values for $N_{\mathrm{thr},i}$ and assess
the deviation of LAT data's and baseline fit data's TS-distributions
energy bin by energy bin. Eventually, we select those ROI masks and
threshold values that produce the statistically most sound masks. 

\emph{Combining the constraints from both data sets.} Despite the
fact that the fitting procedures are adapted individually for each data set,
we can nonetheless derive a combined constraint on the ALPs' signal
strength via a joint-likelihood approach. Eq.~~\ref{eq:TS_stat_reach}
is valid in both cases and the signal templates are generated from
the same input model. Thus, the normalization parameter $A^{\mathrm{ALP}}$
has the same meaning for both data sets. The joint-likelihood that
we utilize within our framework is hence the sum of both weighted
likelihood functions.

\section{Results}
\label{sec:results}

\subsection{Suitable regions of interest}
\label{sec:optimal_roi}

Following the recipes outlined in Sec.~\ref{sec:methodology} to single out a suitable analysis region for both LAT data sets, we present here the final results of this search. We stress again that in the context of this optimization step the IE is represented by the Lorimer I model.

\begin{figure*}[t]
\begin{centering}
\includegraphics[width=0.49\linewidth]{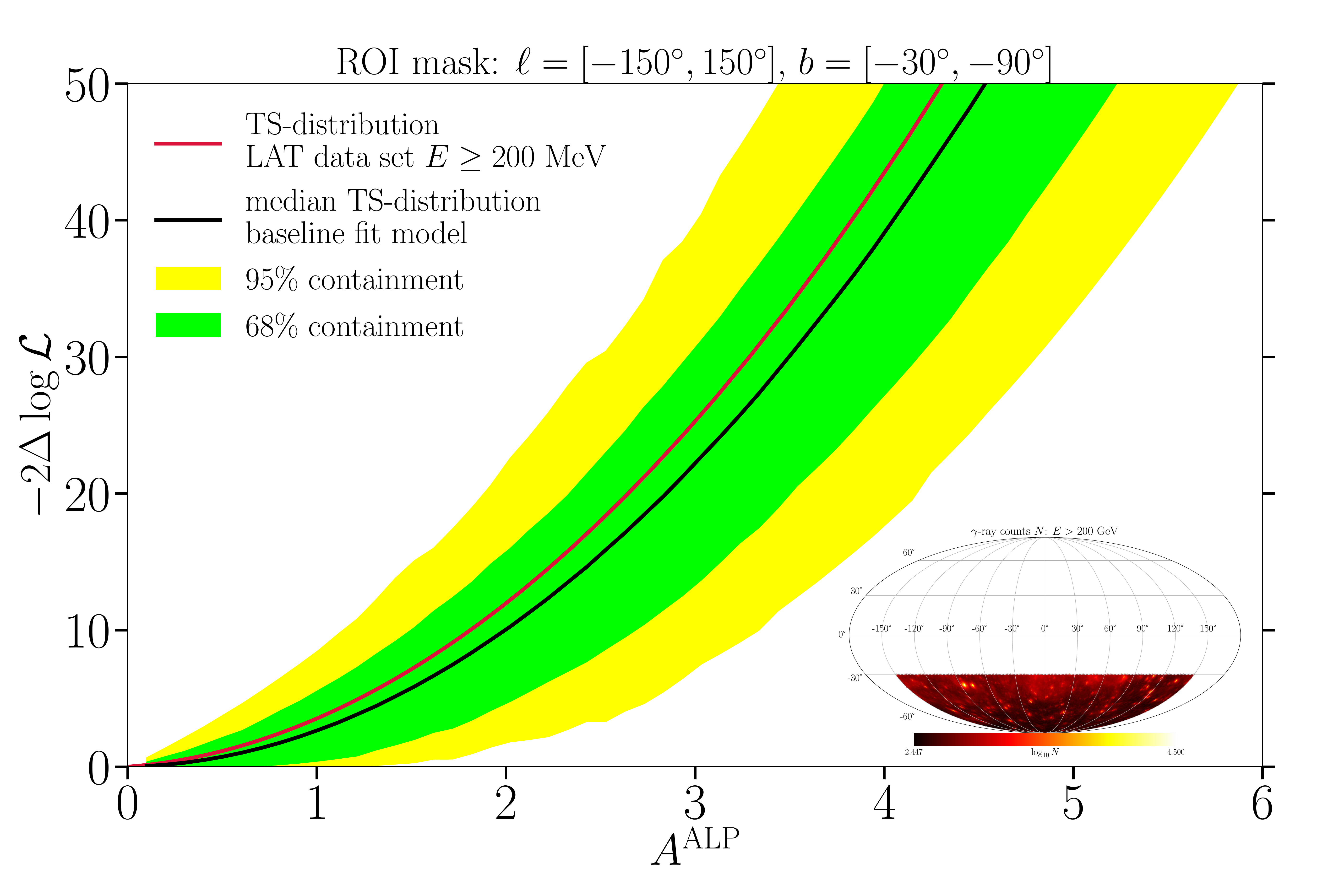}
\includegraphics[width=0.48\linewidth]{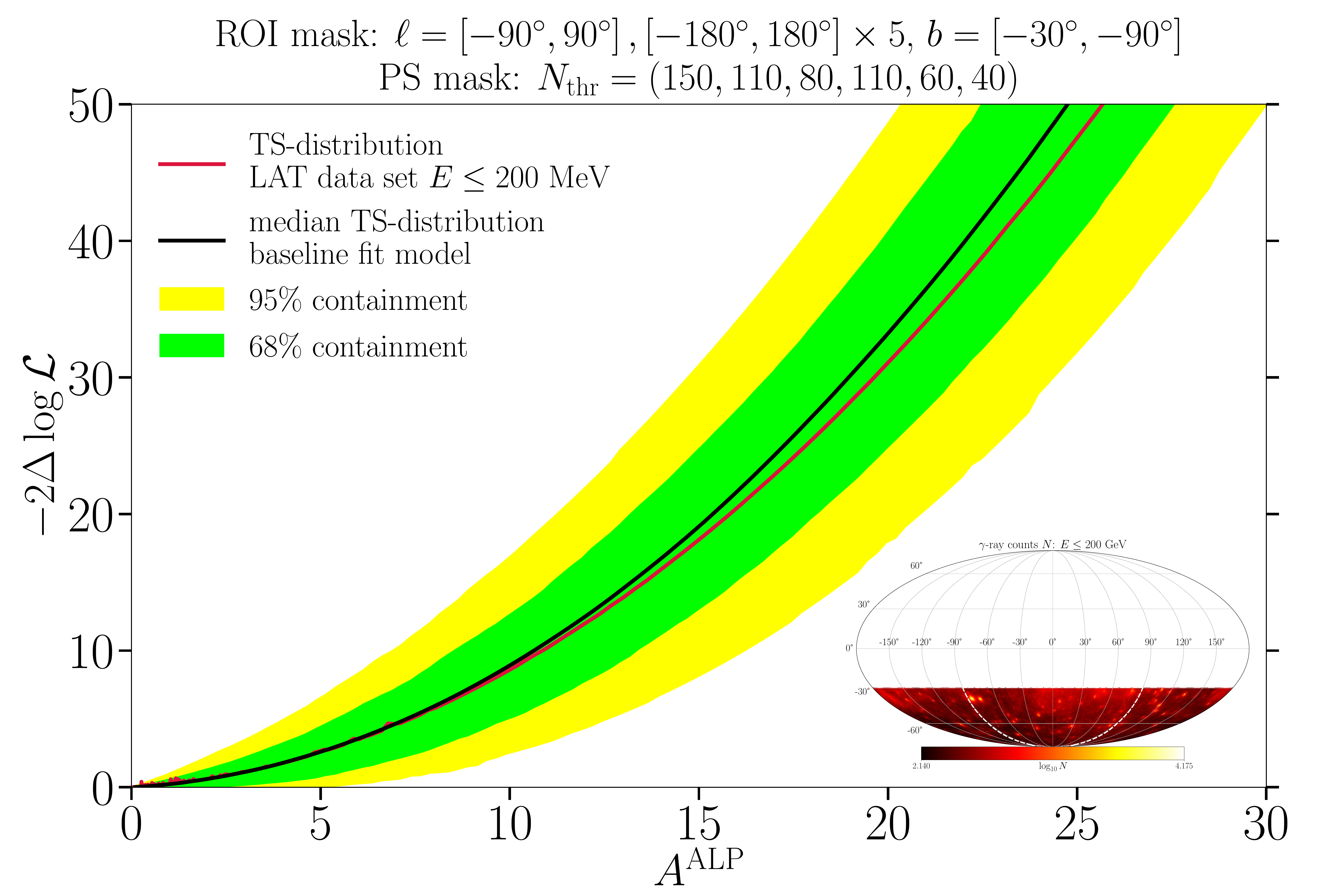}
\par\end{centering}
\caption{(\emph{Left:})  TS-distribution with respect to the baseline fit model (based on Lorimer I; black) and its statistical scatter (green: 68$\%$ containment, yellow: 95$\%$ containment) as well as the TS distribution with respect to the 12-year Fermi-LAT data including energies above 200 MeV (red). (\emph{Right:}) Same as the left panel with respect to the LAT data set of events with $E\leq200$ MeV. The fitting procedure uses six macro energy bins. In the lower right corner of each panel we display the optimized ROI showing the total gamma-ray counts in the respective data set projected on the sky.\label{fig:optimal_ROI}}
\end{figure*}

In Fig.~\ref{fig:optimal_ROI} we display the comparison of the TS-distributions obtained from the LAT data set under study (red) and the baseline fit model (black). The scatter of the TS-distribution is shown as $1\sigma$ and $2\sigma$ containment bands. The left panel of this figure refers to the data set with $E\geq200$ MeV, for which we find the best agreement between real data and model for an ROI with $-90^{\circ}<b<-30^{\circ}, |\ell|\leq150^{\circ}$. The right panel of the same figure shows the situation for the data set below 200 MeV using six macro energy bins. The minimal deviation of LAT data and baseline fit model TS-distribution is ensured by using the 4FGL source mask threshold values $N_{\mathrm{thr}}=\left(150,110,80,110,60,40\right)$ with $\ell_{\mathrm{max}}=180^{\circ}$ for all but the first energy bin where $\ell_{\mathrm{max}}=90^{\circ}$ optimizes the agreement. Again, the Galactic latitude is set to $-90^{\circ}<b<-30^{\circ}$ to reduce the impact of the IE.

\subsection{Upper limits on the ALP parameter space}

After having determined in Sec.~\ref{sec:optimal_roi} the ROI that yields the most statistically sound upper limits
on the ALP signal, we are able to set upper limits on the normalization
parameter $G_a$ of the signal template. Before that,
we have checked that the selected parts of the sky do not contain
a significant fraction of the ALP signal that would warrant a detection.
We ``unblind'' our previous fitting routine by inserting the true
signal template with the gamma-ray flux spectrum induced by ALPs from
core-collapse SNe, Sec.~\ref{sec:DSNALPB_theory}; hence, the re-introduction of the normalization parameter $G_a$. 

In what follows, we consider and utilize a benchmark case of the DSNALPB gamma-ray spectrum to illustrate the upper limits on such a large-scale gamma-ray emission component. This benchmark model is defined by the following properties:
\begin{enumerate}
    \item $f_{\rm fail-CC} = 20\%$,
    \item Salpeter IMF,
    \item fiducial SNR description (see Tab.~\ref{tab:RSFparameters}).
\end{enumerate}
The uncertainty on the reported DSNALPB upper limits arising from varying these benchmark choices is discussed in Sec.~\ref{sec:discussion}. Therein, we also report the impact of altering the astrophysical surroundings of the Milky Way, that is, the employed IEM and GMF model.

%We consider the DNALPB gamma-ray spectrum, propagating the uncertainty band in Fig.~\ref{fig:dsalpfl}. \JJC{Fix label}

We consider ALPs coupled only to photons. 
In this case, 
the upper limit on $G_a$ translates
into an upper limit on the photon-ALP coupling strength via $g_{a\gamma}=\sqrt[4]{G_a} \,g_{a\gamma}^{\mathrm{ref}}$, where $g_{a\gamma}^{\mathrm{ref}} = 5\times10^{-12}\;\mathrm{GeV}^{-1}$ refers to the reference value of the coupling at which spectrum and ALP-photon conversion probability in the Milky Way have been calculated to obtained the ALP template.
%\JJC{Add reference value}

\begin{figure*}[t]
\begin{centering}
\includegraphics[width=0.75\linewidth]{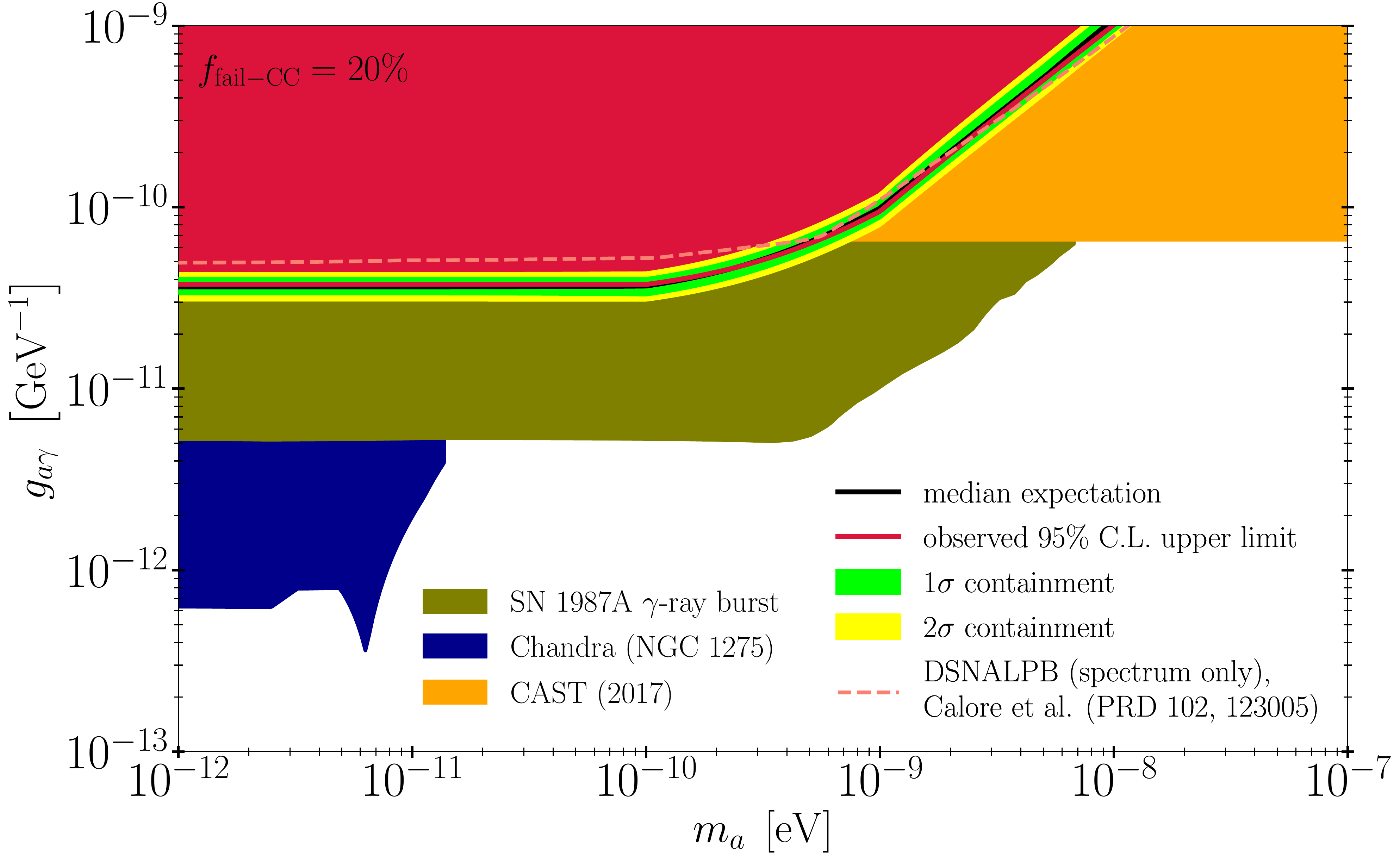}
\par\end{centering}
\caption{95$\%$ CL upper limits (solid red) on the ALP-photon coupling constant $g_{a\gamma}$ assuming the benchmark DSNALPB scenario and an ALP coupling exclusively to photons as well as the `Jansson12c' \cite{2016A&A...596A.103P} model of the Milky Way's GMF. The filled red region illustrates the ALP parameter space excluded by this upper limit. The displayed green (yellow) band reflects the expected $1\sigma\, (2\sigma)$ statistical scatter of the upper limits based on 250 Poisson realizations of the ``mock data'' obtained via the baseline fit of the gamma-ray sky. The solid black line represents the median upper limit obtained from these fits to mock data. To highlight the improvement on the upper limits set with the analysis in Paper I, which is solely based on the expected spectral shape of the DSNALPB gamma-ray flux, we show this result as a dashed, light-red line. Our results are complemented by independent astrophysical and helioscope bounds on the ALP-photon coupling strength from CAST \cite{CAST:2017uph}, Chandra observations of NGC 1275 \cite{Reynolds:2019uqt} as well as the non-observation of a gamma-ray burst following SN 1987A \cite{Payez:2014xsa}.  \label{fig:limits_ALP_Jansson}}
\end{figure*}

In Fig.~\ref{fig:limits_ALP_Jansson} we show the observed 95$\%$ CL upper limits (solid red line) on our benchmark DSNALPB scenario together with the expected statistical scatter ($68\%$ containment: green; $95\%$ containment: yellow) of the upper limits according to 250 Poisson realizations of the baseline gamma-ray sky model (cf.~Sec.~\ref{sec:methodology} for its derivation) whereas the solid black line denotes the median upper limit with respect to this baseline data set. The basis for the baseline sky model and all derived upper limits presented here are the ALP signal morphology due to the Jansson model \cite{2012ApJ...757...14J} of the Milky Way's GMF and the iteratively optimized IEM Lorimer I. \JJ{We also} confront the upper limits obtained in this analysis with existing limits on the ALP parameter space.

%due to the different scenarios of the ratio of failed to successful CC SNe, i.e.~this band is the one-to-one translation of the uncertainty band in Fig.~\ref{fig:dsalpfl} \JJC{Fix label} to an uncertainty on the upper limits on $g_{a{\gamma}}$.

%In Sec.~XXX we compare the dependence of our results on different characterizations of the
%magnetic field while in Sec.~XXX we investigate the robustness of these constraints under variations of the IEM.

In this particular setting, we find an improvement of the upper limit on $g_{a\gamma}$
regarding our previous analysis in Paper I that was solely
based on the spectral shape of the ALP-induced gamma-ray flux (and neglecting the effect of gravitational energy-redshift as well as the formation of alpha particles during a CC SN). 
Specifically, we obtain $g_{a\gamma}\lesssim 3.76\times 10^{-11}$ GeV$^{-1}$
for ALP masses $m_{a}\ll10^{-11}$ eV at $95\%$ CL.

\bigskip

%\sout{Finally, let us note that the whole procedure can easily be applied also to other production  mechanisms. For example, if we also consider a coupling between ALPs and 
%nucleons, the limit on the ALP-photon coupling is significantly improved. 
%In our analysis set-up -- assuming $g_{aN}=10^{-9}$ and very naively the same factor of improvement found in \NEW{Paper I} -- we obtain $g_{a\gamma}\lesssim3.4\times 10^{-13}$
%GeV$^{-1}$ for ALP masses $m_{a}\ll10^{-11}$ eV regarding the benchmark DSNALPB scenario.}
%\JJC{Does this include information from the improved simulations, i.e. using also the different mass SNe?}

\section{Discussion}
\label{sec:discussion}

% Magnetic field uncertainties
% Planck paper: https://arxiv.org/pdf/1601.00546.pdf

This section is dedicated to a discussion of the sources of systematic uncertainties on the DSNALPB upper limits reported in Sec.~\ref{sec:results}. These uncertainties arise by varying the benchmark scenario decisions as well as the description of the astrophysical surroundings in the Milky Way.

While a number of dedicated explorations of particular sources of uncertainty regarding their impact on the ALP-photon coupling upper limits are given in App.~\ref{app:B}, we provide below in Tab.~\ref{tab:upperlim_uncertainty} a summary of the induced systematic uncertainty for the ``massless'' ALP case $m_a \ll 10^{-11}$ eV while always referring to the benchmark DSNALPB scenario as reference point.

\begin{table}[!h]
    \centering
    \begin{tabular}{l c c}
         \hline
         source of uncertainty \hspace*{5pt} & absolute $\left[10^{-11}\;\mathrm{GeV}^{-1}\right]$\hspace*{5pt} & relative $\left[\%\right]$\hspace*{5pt}\\
         \hline
         \hline
         %statistical ($1\sigma$) & $\left[2.89, 3.32\right]$ & $13.1$\\
         %statistical ($1\sigma$) & $\left[3.49, 4.14\right]$ & $17.3$\\
         %$f_{\mathrm{fail-CC}}$ & $\left[2.52, 3.97\right]$ & $44.2$\\
         $f_{\mathrm{fail-CC}}$ & $\left[2.81, 4.73\right]$ & $51.1$\\
         %IMF & $\left[3.29, 3.53\right]$ & $7.3$\\
         IMF & $\left[3.76, 4.03\right]$ & $7.2$\\
         %SNR & $\left[3.15, 3.49\right]$ & $10.3$\\
         SNR & $\left[3.59, 3.98\right]$ & $10.4$\\
         %IEM & $\left[2.83, 3.29\right]$ & $14.0$\\
         IEM & $\left[3.24, 3.76\right]$ & $13.8$\\
         %GMF model & $\left[3.29, 4.57\right]$ & $38.9$\\
         GMF model & $\left[3.76, 5.22\right]$ & $38.8$\\
         \hline
         total & $\left[2.38, 7.04\right]$ & $124$\\
         \hline
    \end{tabular}
    \caption{Induced uncertainty on the ALP-photon coupling constant $g_{a\gamma}$ with respect to varying the conditions and properties assumed in the benchmark DSNALPB scenario and $m_a \ll 10^{-11}$ eV. %\sout{For each source of uncertainty, we report the absolute uncertainty range of the derived respective upper limits as well as the relative uncertainty taken with respect to the nominal value of the upper limit for the benchmark case. To be more explicit, the quoted statistical uncertainty range corresponds to the green band in Fig.~\ref{fig:limits_ALP_Jansson}; the $f_{\mathrm{fail-CC}}$ uncertainty range reflects the grey band in Fig.~\ref{fig:dsalprange}; the IMF uncertainty arises from the two alternative initial mass functions Kroupa and BG (see Sec.~\ref{sec:DSNALPB_theory}); the SNR uncertainty uses the remaining parametrizations in Tab.~\ref{tab:RSFparameters}, the IEM uncertainty range uses the five different models introduced in Sec.~\ref{sec:methodology} and the GMF model uncertainty reflects the change from the Jansson \& Farrar prescription to the Pshirkov model.}\FC{I moved this in the text} 
    The last row indicates the total systematic uncertainty range when all sources of uncertainty are combined to form a most optimistic and most pessimistic scenario.
    \label{tab:upperlim_uncertainty}}
\end{table}

For each source of uncertainty -- listed in the first column -- we report in the second column the associated absolute uncertainty range of the derived $95\%$ CL upper limits. The stated absolute range quantifies the minimal and maximal constraint that we find by varying the respective quantity within its uncertainty range, while keeping all other quantities fixed to their values attained in the benchmark case. These boundaries do not need to be symmetric around the benchmark upper limit depending on the source of uncertainty. For example, we only consider one alternative GMF model so that the reported interval refers to the numbers obtained with respect to either the Jansson \& Farrar model or the Pshirkov model. 
The third column of Tab.~\ref{tab:upperlim_uncertainty} contains the relative uncertainty taken with respect to the nominal value of the upper limit for the benchmark case. This means, we take the difference between the lower and upper boundary in the first column and divide it by the benchmark upper limit.

To be more explicit regarding the origin the table's content, 
%the quoted statistical uncertainty range corresponds to the green band in Fig.~\ref{fig:limits_ALP_Jansson}; 
the $f_{\mathrm{fail-CC}}$ uncertainty range reflects the grey band in Fig.~\ref{fig:dsalprange}; the IMF uncertainty arises from the two alternative initial mass functions Kroupa and BG (see Sec.~\ref{sec:DSNALPB_theory}); the SNR uncertainty uses the remaining parametrizations in Tab.~\ref{tab:RSFparameters}, the IEM uncertainty range uses the five different models introduced in Sec.~\ref{sec:methodology} and the GMF model uncertainty reflects the change from the Jansson \& Farrar prescription to the Pshirkov model.
This assessment of the systematic uncertainties of our upper bounds singles out the unknown fraction of failed CC SNe, as well as the strength and structure of the Milky Way's GMF as the most significant drivers of uncertainty, contributing an error of about 51\% and 39\% respectively.\\
On the other hand, the uncertainty related to the IMF, SNR parametrization, and IEM account for a $\sim$ 10\% relative error.
%, which is at the same level as the statistical uncertainty.}
%

%The values in Tab.~\ref{tab:upperlim_uncertainty} are a fairly generous estimate of the corresponding uncertainty since they have been derived by varying only a single source of uncertainty at a time.

\begin{figure*}[t]
\begin{centering}
\includegraphics[width=0.75\linewidth]{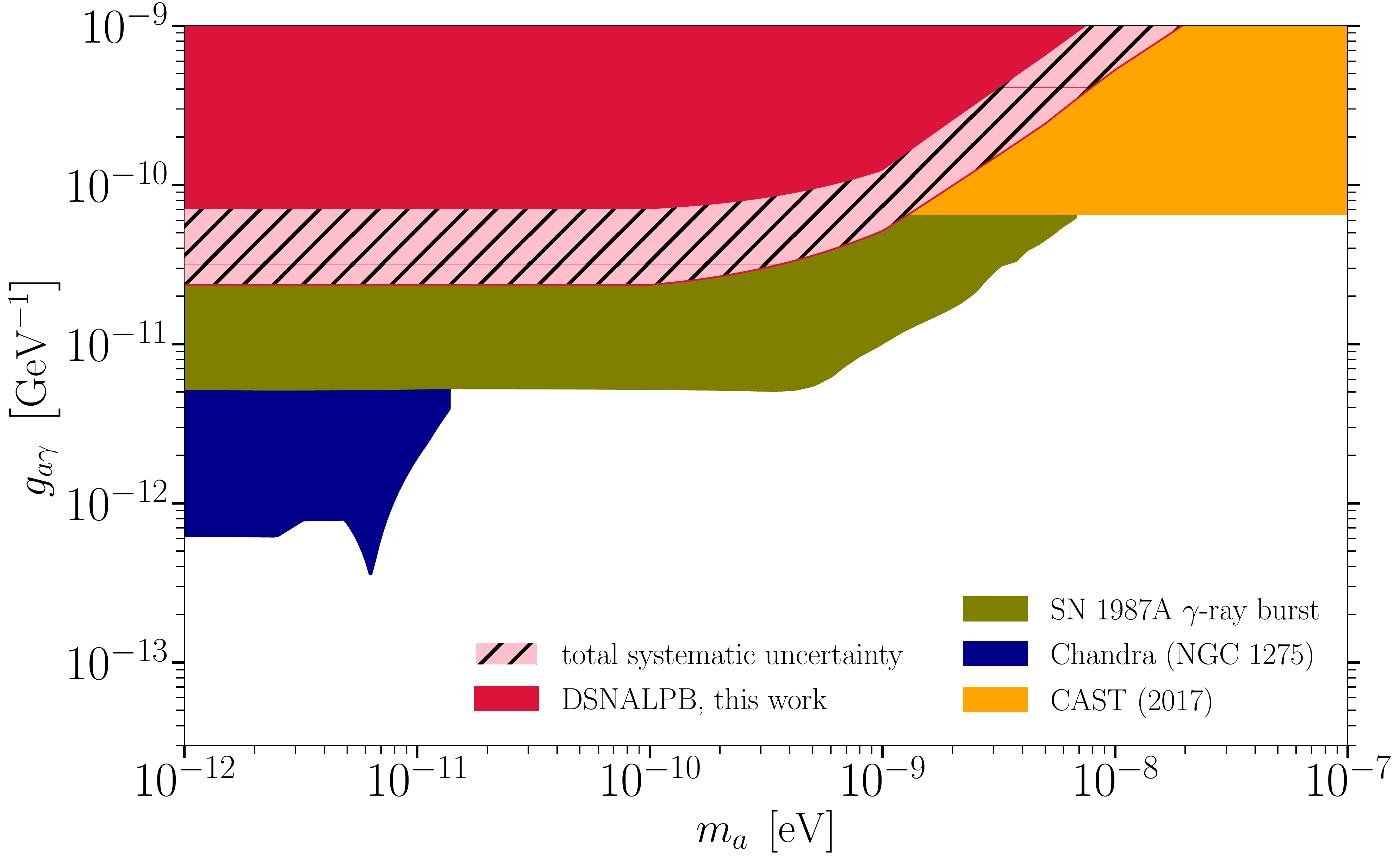}
\par\end{centering}
\caption{95$\%$ CL upper limits (red band) on the ALP-photon coupling constant $g_{a\gamma}$ assuming a coupling exclusively to photons. The displayed pink hatched band reflects the total systematic uncertainty on the DSNALPB gamma-ray spectrum caused by combining all sources of uncertainty considered in this analysis (see Tab.~\ref{tab:upperlim_uncertainty}) to form a most optimistic and a most pessimistic scenario. Our results are complemented by independent astrophysical and helioscope bounds on the ALP-photon coupling strength from CAST \cite{CAST:2017uph}, Chandra observations of NGC 1275 \cite{Reynolds:2019uqt} as well as the non-observation of a gamma-ray burst following SN 1987A \cite{Payez:2014xsa}.  \label{fig:full_syst_uncertainty}}
\end{figure*}

The values in the first lines of Tab.~\ref{tab:upperlim_uncertainty} provide an estimate of the uncertainty due to individual inputs since have been derived by varying only a single source of uncertainty at a time.
To get an impression of the overall uncertainty we consider most optimistic and pessimistic scenarios that lead to the best or worst possible upper limits on the DSNALPB. The resulting systematic uncertainty band is displayed in Fig.~\ref{fig:full_syst_uncertainty}.
With respect to these most optimistic and pessimistic scenarios, the systematic uncertainties may allow the $95\%$ CL upper limit on the DSNALPB to be placed between $g_{a\gamma} \lesssim \left[2.38, 7.04\right]\times10^{-11}\;\mathrm{GeV}^{-1}$ in the case of massless ALPs.

Further details on the uncertainties can be found in App.~\ref{app:B}.

\section{Conclusions}
\label{sec:conclusion}
In this work, we carried out a comprehensive analysis of the gamma-ray diffuse signal 
induced by axion-like particles (ALPs) produced by all past cosmic core-collaps supernovae (CC SNe) and converted into high-energy photons
when experiencing the magnetic field of the Galaxy.

We presented a refined calculation of the ALPs flux from extragalactic SNe: 
we go beyond the simple approximation that the ALPs flux is independent on the SN mass progenitor
by modeling the ALPs signal from different state-of-the-art SN models with 
progenitor masses between 8.8 and 70 $M_\odot$.
Moreover, we accounted for the possibility that not all CC SNe lead to 
successful explosions, by quantifying the fraction of failed CC SN explosions
and building the corresponding model for the ALPs signal upon two simulations of 
failed CC SN explosions. 

We explored four different scenarios, each of them characterized 
by a different fraction of failed CC SN explosions, allowing us to quantify the uncertainty 
due to failed CC SN explosions.
The calculation of the ALPs flux from all past cosmic SNe accounts also
for uncertainties related to the cosmic SN rate.

Using this new model for the diffuse supernova ALP background (DSNALPB) gamma-ray flux, we run a systematic analysis 
of 12 years of data collected by \Fermi-LAT with the aim of setting robust upper limits 
on the ALPs parameter space. 
For the first time in the context of ALPs searches, we performed a template-based 
gamma-ray analysis to fully exploit the spatial features of the ALPs signal.
The flux from the DSNALPB being peaked at about 25~MeV, we exploited the full LAT data sets
by developing an optimized low-energy ($E \lesssim 200$ MeV) analysis.
Besides, we optimized the IEMs in a data-driven way and limit the impact of the IE mis-modeling 
on the final limits -- which, indeed, are only mildly affected by changing the IEM.
We also selected the ROI in order to be able to set statistically sound upper 
limit on the signal model. 

Our final limits slightly improve the CAST bound (in the low mass region). However, they are still about a factor of six (regarding our baseline scenario) above the 
SN1987A gamma-ray burst limit. It is nevertheless a valuable confirmation, as they do not depend on a single event.
More importantly, we quantified for the first time the width of the uncertainty band of the DSNALPB limit,
which turns out to be less than a factor of three and dominated by the uncertainty on the 
fraction of failed CC SN explosions.
A significant improvement on our bound would be therefore reached exploiting the synergies with the detection of the future diffuse supernova neutrino background (DSNB). Indeed, as pointed out in Ref.~\cite{Moller:2018kpn}, a combined detection of the DSNB in the next-generation neutrino detectors
%\JJ{\sout{, i.e. Hyper-Kamiokande enriched with Gadolinium, JUNO, and DUNE, after 20 years of data taking}}
will be sensitive to the local supernova rate at a $\sim 33$ \% level, and will give  an uncertainty on the fraction of supernovae that form black holes that will be at most $\sim 0.4$.  Consequently, the uncertainty on the DSNALPB flux would be significantly reduced.

Uncertainties on the IEM are sub-dominant, while those on the GMF
remain an important source of systematic uncertainties for ALPs searches. 
In this respect, we stress that only the transversal component to the ALPs' propagation is relevant for the conversion. Diffuse synchrotron in radio- and microwaves and thermal dust emission are crucial for constraining 
GMF models perpendicular to the line-of-sight, and complement each other. 
Improvements on our description of the GMF are expected by new radio- and microwave surveys (e.g.~SKA, QUIJOTE), as well as from the synergy between GAIA and {\it Planck} through a detailed mapping of the dust distribution
via extinction. SKA will also allow the scientific community to make a leap forward in the number of pulsars known in the Galaxy (and therefore in Faraday rotation data), and to refine our model for electron density and the parallel magnetic field component.
A better comprehension of the Galactic cosmic-ray population from AMS-02 future measurements and gamma-ray telescopes, joint with synchrotron maps, will also help us constraining the GMF ordering.
We refer the reader to~\cite{Jaffe:2019iuk} for a more detailed discussion and overview.

%\FC{Don't like very much this concluding sentence. Perhaps we can simply remove it}
%\sout{Although the numerical values of our limits change only by a modest amount compared to previous
%results, we believe that the present work
%offers a uniquely coherent and robust analysis of the DSNALPB signal with \Fermi-LAT data.}
To conclude, we have presented here a first, systematic, analysis of the ALPs diffuse background from CC SNe with gamma-ray data, leveraging on the unique sensitivity of the \Fermi-LAT.
%{\bf let us remove it }

%{\bf is it possible to put the final figurs of the Appendix in a single page before the references?}

\begin{acknowledgments}
We warmly thank Giuseppe Lucente for useful discussions during the development of this project.
We would like to acknowledge the anonymous referee of this manuscript for the helpful comments which 
contributed improving the quality of the scientific output.
The work of P.C. is partially supported by the European Research Council under Grant No. 742104 and by the Swedish Research Council (VR) under grants  2018-03641 and 2019-02337.
The work of P.C. and 
A.M. is partially supported by the Italian Istituto Nazionale di Fisica Nucleare (INFN) through the ``Theoretical Astroparticle Physics'' project
and by the research grant number 2017W4HA7S
``NAT-NET: Neutrino and Astroparticle Theory Network'' under the program
PRIN 2017 funded by the Italian Ministero dell'Universit\`a e della
Ricerca (MUR). 
The work of C.E. is supported by the "Agence Nationale de la Recherche”, grant n. ANR-19-CE31-0005-01 (PI: F. Calore).
T.F. acknowledges support from the Polish National Science Center (NCN) under Grant No. 2020/37/B/ST9/00691.
 K.K. acknowledges support from Research Institute of Stellar Explosive Phenomena (REISEP) at Fukuoka University and also from the Ministry of Education, Science and Culture of Japan (MEXT, No.JP17H06357)
and JICFuS as “Program for Promoting researches on the Supercomputer Fugaku” (Toward a unified view of 
the universe: from large scale structures to planets, JPMXP1020200109). 
Numerical computations of T.K. were carried out on Cray XC50 at CfCA, NAOJ and on Sakura and Raven at Max Planck Computing and Data Facility.
The work of M.G.~is partially supported by a grant provided by the Fulbright U.S.~Scholar Program and by a grant from the Fundación Bancaria Ibercaja y Fundación CAI. 
M.G.~thanks the Departamento de Física Teórica and the Centro de Astropartículas y Física de Altas Energías (CAPA) of the Universidad de Zaragoza for hospitality during the completion of this work.

\end{acknowledgments}

\appendix

	\section{Details on the calculation of the ALP spectrum\footnote{Once more we would like to thank the anonymous referee for raising our awareness of these effects.}}
	
\label{app:A}	
	
	\subsection{Impact of the alpha particles}

\begin{figure}[t!]
		\vspace{0.cm}
		\includegraphics[width=0.95\columnwidth]{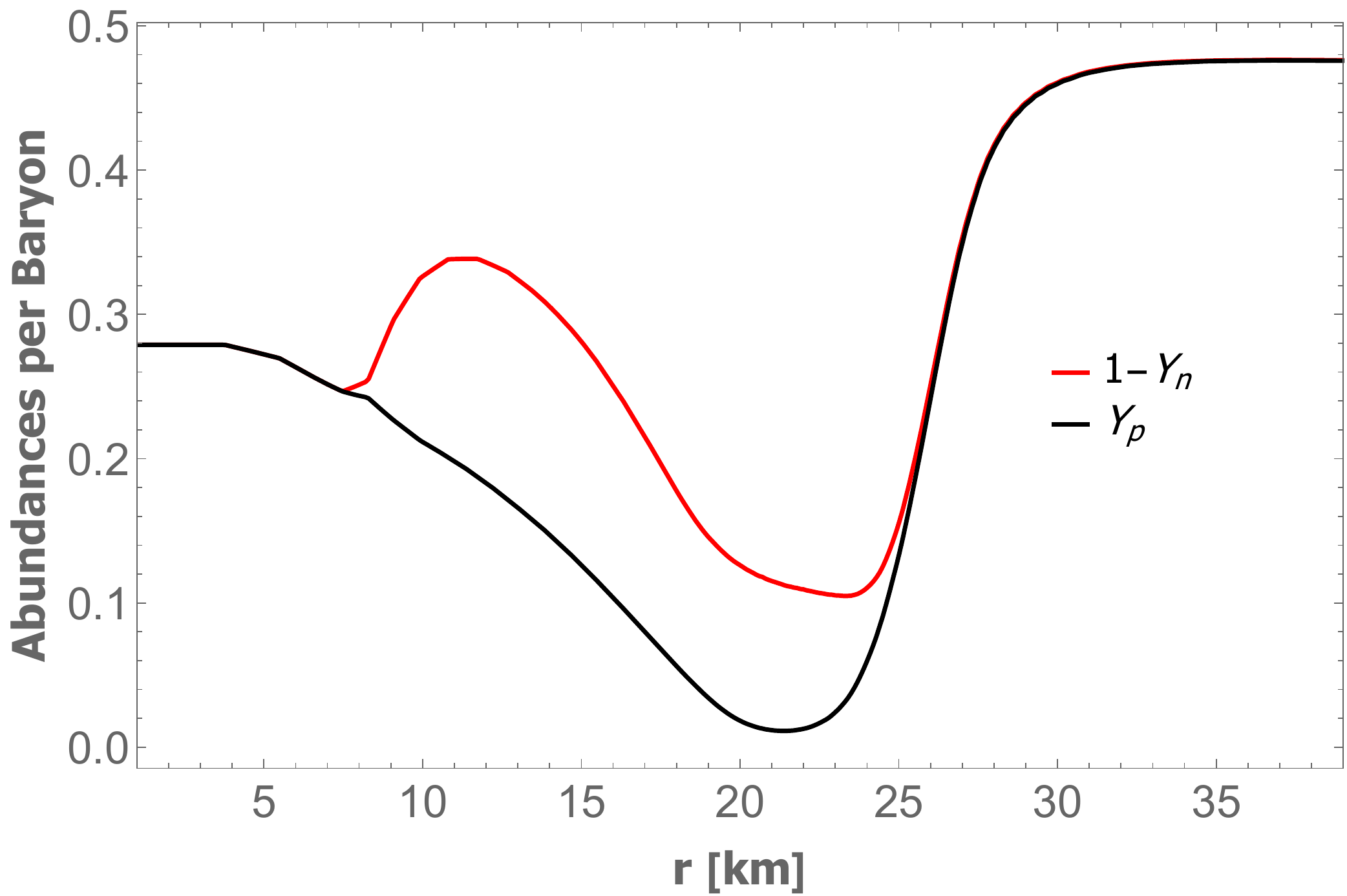}
		\caption{Charged particles abundances a $t_{\textrm{pb}}=1$ s for the model $M=25\,M_\odot$.}
		\label{fig:YnYp}
	\end{figure}

\begin{figure}[t!]
		\vspace{0.cm}
		\includegraphics[width=0.95\columnwidth]{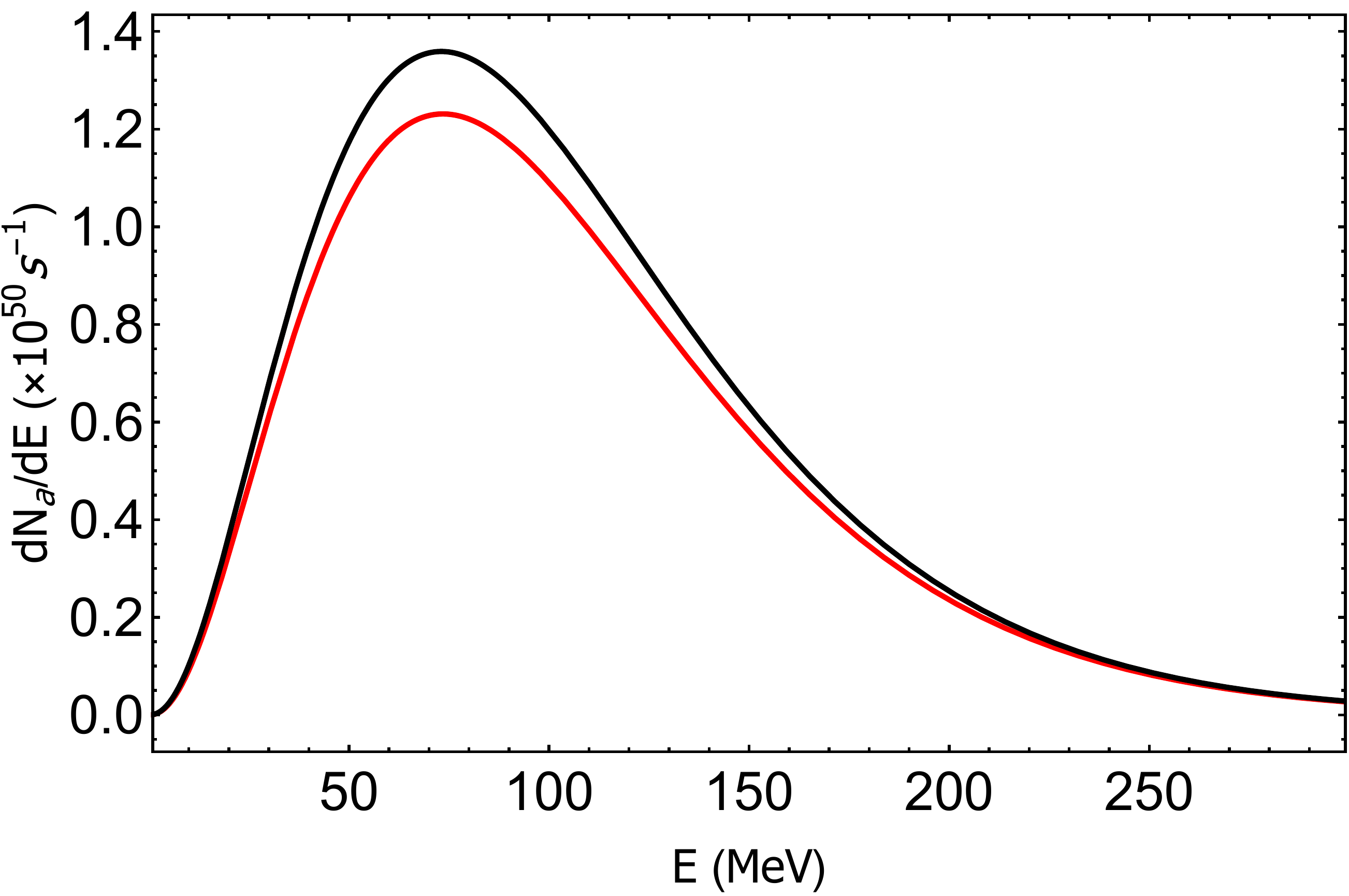}
		\caption{Comparison beetween produced SN ALP number flux as a function of energy taking into account  the Primakoff production only  from protons (red line) and including also the production    from $\alpha$ particles (black line). The chosen model is the $M=25\,M_\odot$. 
			We assume 
			%\sout{$m_a \ll 10^{-11}$~eV} 
			$g_{a\gamma}=10^{-11}\GeV^{-1}$.
			}
		\label{fig:dNadEalpha}
	\end{figure}

Usually the ALP Primakoff production in SN has been characterized 
including only the contributions from protons. 
However, as recently pointed out in~\cite{Caputo:2021rux}
the contribution of alpha particles in the SN core might be non negligible.
Indeed, we confirm that also for the SN models we use, 
there is a sizable  gap between the proton abundance 
$Y_p$ and  $1-Y_n$, as shown in 
 Fig.~\ref{fig:YnYp}, that we can assume to be filled by alpha particles.\\
In order to evaluate the effect of these particles on the ALP production, a reasonable choice according to~\cite{Caputo:2021rux} is to correct the inverse Debye screening length $\kappa$ described by
\begin{equation}
    \kappa^2 =\frac{4\pi \alpha \hat{n}}{T}\;, 
\end{equation}
where $\hat{n}=\sum_j Z_j^2 n_j=\hat{Y}n_{\rm B}$, where
$\hat{Y}$ is the effective charge per nucleon.
 If all nuclei heavier than protons were realized as $\alpha$ particles, we would have $X_\alpha+X_n+X_p=1$, where $X_j$ represents the mass fraction for the particle $j$. In this framework $\hat{Y}=Y_p+4X_\alpha/4=Y_p + X_\alpha$.
 The difference in the SN energy spectrum can be observed for the $25 M_{\odot}$ SN progenitor is shown in Fig.~\ref{fig:dNadEalpha}. We find that the inclusion of alpha particles produces an enhancement of $\sim 15 \%$
 of the ALP flux.


	\subsection{Gravitational energy-redshift
	%\footnote{\NEW{We again would like to thank the anonymous referee for pointing us towards this significant effect.}
	}\label{app:GR}

The ALPs emission is affected by the gravitational field of the neutron star\footnote{During the final stages of completion of the improved version of this manuscript a new paper~\cite{Caputo:2022mah} appeared which also discusses this effect.}, in particular time dilation, trajectory bending and the red-shifting of the energy.
In this appendix we discuss the implementation of these gravitational effects in our analysis.\footnote{We are very grateful to Giuseppe Lucente for sharing his thoughts and his notes on the subject.}

Let us start with a couple of general comments.
We are calculating the time integrated production in the local reference frame. As we are not interested in the time dependence of the signal, particle number conservation ensures that we have the correct number of ALPs also outside the supernova. As we are considering an isotropic flux, trajectory bending can be ignored. The most important effect for us is the red-shift of the energy, because it directly affects the spectrum which in turn determines the sensitivity of Fermi-LAT.

All SN simulations discussed in the present manuscript are based on general relativistic neutrino radiation hydrodynamics~\cite{Liebendoerfer:2004,KurodaT18,Kuroda:2021}, i.e. the metric functions are obtained through direct numerical integration of the Einstein field equations for a given line element, $ds^{2}= g_{\alpha \beta} dx^\alpha dx^\beta$. The zeroth component, known as the lapse function, $g_{00}=-\exp\{2\Phi\}$, determines the gravitational red(blue) shifting of the axion energy as follows,
%.......
\begin{equation}
    E=E^*(x)\,\textrm{exp}(\Phi(x))\;,
    \label{eq:redsh}
\end{equation}
%.........
with the lapse function being evaluated locally at the PNS interior, depending on the choice of the coordinate system $\{x^\alpha\}$, relating the ALP energy $E$ measured by an observer at infinity with the local ALP energy $E^*(x)$. 
Similarly, for a local observer, time dilation must be taken into account as follows
%............
\begin{equation}
dt= dt^*(x)  \,\ \textrm{exp}(\Phi(x))\;,   
\end{equation}
%.............
where the $dt^*(x)$ refers to the local observer time
at $x$, while $dt$ refers to the simulation time corresponding
to that of a distant observer.

%%%%%%%%%%%%%%%%%%%
\begin{figure}[t!]
\vspace{0.cm}
\includegraphics[width=0.95\columnwidth]{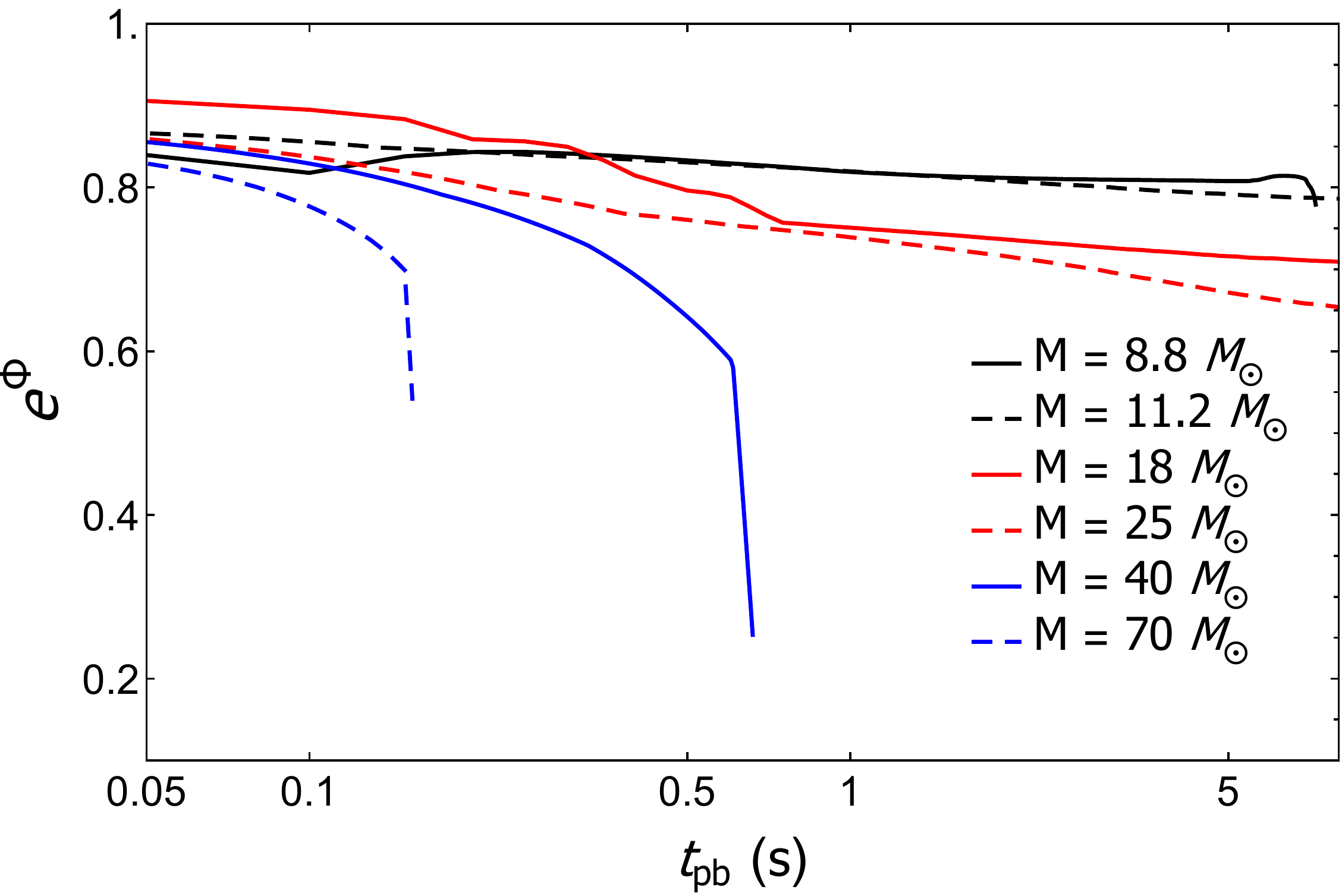}
\caption{Post-bounce evolution of the lapse function $e^{\Phi}$ sampled at a distance from the center of $r=5$~km for different SN simulations under investigation.}
\label{fig:lapse}
\end{figure}
%%%%%%%%%%%%%%%%

%We assume  spherical symmetry and we neglect rotation and \JJ{the} magnetic field in the SN core. With these approximations, we can describe adequately the region outside the neutron star as the  Schwarzschild geometry with metric	 componente $g_{\alpha \beta}$ defined through the line element $ds^{2}= g_{\alpha \beta} dx^\alpha dx^\beta$~\cite{Misner:1973prb}
%..........	
%\begin{equation}
%    ds^{2} = -e^{2 \Phi (r)}dt^{2}+2e^{2 \Lambda (r)}dr^{2}+r^{2}d\Omega^{2}\;,
%    \label{Eq.metric1}
%\end{equation}
%.................
%where  $d\Omega^{2}= (d\theta^{2} + \sin^{2} \theta d\phi^{2})$, and ${x^{\alpha}}=(t,r,\theta, \phi)$ are the  Schwarzschild coordinates. Here $\Phi (r)$ and $\Lambda (r)$ are the usual radial coordinate-dependent Schwarzschild metric functions.
%
%The observed ALP energy $E$ measured at infinity is related to  the ALP emission energy $E^\ast(r)$ measured by a locally inertial observer at rest at the emission radius $r$ in the SN by~\cite{Fuller:1996kt}
%.......
%\begin{equation}
%    E=E^{*} \textrm{exp}(\Phi (r))\;,
%    \label{eq:redsh}
%\end{equation}
%.........
%{\bf EXPLAIN how $\phi(r)$ is calculated}

In Figure~\ref{fig:lapse}, we show the time evolution of the lapse function $e^{\Phi}$ for the different SN progenitors at a distance from the center of $r=5$~km. We note that for exploding SNe this factor decreases monotonically in time, the effect being larger for higher progenitor masses, and ranging between 0.7--0.8. For failed SN collapsing to black-holes, the gravitational effect is larger, i.e. the lapse function is dropping to $\sim 0.5$ for the $70$ $M_{\odot}$ progenitor already shortly after core bounce and below $0.01$ when the apparent horizon appears at $t_{\rm pb}\sim0.155$~s \citep{Kuroda:2021}.

The ALP Primakoff production rate of 
Eq.~(\ref{eq:axprod}) in terms of the local quantities  reads
%{\bf [I WRITE EXPLICITLY $t^{*}$]}
%...........................
\begin{equation}
    \frac{d{n}_{\rm a}}{dE^{*}dt^{*}}=\frac{g_{\rm a\gamma}^{2}\xi^{2} T^{3} (E^{*})^{3}}{8 \pi^{3}(e^{E^{*}/T-1})}\biggl[ \frac{\xi^{2}T^{2}}{(E^{*})^{2}} \textrm{ln}\biggl(1+\frac{(E^{*})^{2}}{\xi^{2}T^{2}}\biggr) \biggr].
\end{equation}
%......................
%\sout{expressed in terms of local measured energy $E^{*}(r)$,
%as to be redshifted according to Eq.~(\ref{eq:redsh})}
%{\bf remove equation}
%......................
%\begin{equation}
%    \frac{d\dot{n}_{\rm a}}{dE}=\frac{d\dot{n}_{\rm a}}{dE^{*}}\frac{dE^{*}}{dE}=\frac{d\dot{n}_{\rm a}(E \cdot \textrm{exp}(-\Phi (r)))}{dE^{*}} \textrm{exp}(-\Phi (r))\;.
%    \label{eq:EPhi}
%\end{equation}
%...............................
Since $dE^{*} dt^{*}= 
dE \,\ dt$, the redshifted
time-integrated
ALP spectrum at infinity is given by
%.........................
\begin{eqnarray}
    \frac{dN_{\rm a}}{dE}
 &=& \int d^{3}r dt \frac{d{n}_{\rm a}}{dEdt} \nonumber \\
 &=&
 \int d^{3}r dt^{*} \frac{d{n}_{\rm a}}{dE^{*}dt^{*}} \textrm{exp}(-\Phi (r)) \,\ .
\end{eqnarray}
%......................................
%\sout{where  we do not explicitly consider the time redshift
%{\bf in the time integral. 
%Indeed, since the integral is extended over all the duration of the burst, due to the conservation of total number of emitted particles, it is not relevant the  time reference frame.
%}
%}

As shown in Sec.~III, the ALP spectrum can be fitted 
by the following functional form
%...................................
\begin{equation}
		\frac{dN_a}{dE} = C \left(\frac{g_{a\gamma}}{10^{-11}\textrm{GeV}^{-1}}\right)^{2}
		\left(\frac{E}{E_0}\right)^\beta \exp\left( -\frac{(\beta + 1) E}{E_0}\right) \,.
		\label{eq:time-int-spec1}
	\end{equation}
%......................

In Table~\ref{tab:spectfit_nored},	% and~\ref{tab:spectfit_red} 
we compare the fitting parameters of Eq.~(\ref{eq:time-int-spec1}) without and with gravitational energy-redshift, respectively for different progenitor masses. 
We see that the effect of gravitational energy-redsfhit 
is to reduce the average energy of the spectrum $E_0$ and increase the normalization parameter $c$ to compensate the drop in $E_0$.  The effect of drop of the energy increases monotonically in function of the SN progenitor mass, ranging from $\sim 20 \%$ for 
8.8 $M_{\odot}$ progenitor to $ 360 \%$ for 
70 $M_{\odot}$. Indeed, increasing the progenitor mass 
we increase the gravitational potential, especially for the high-mass progenitor cases ending into a black-hole. The effect on the $C$ parameter is milder, the increase 
being at most $\sim 30 \%$.
The factor $\beta$ being given by
%................
\begin{equation}
\frac{\langle E^2 \rangle-\langle E \rangle^2}{\langle E \rangle^2}= \frac{1}{1+\beta} \,\ ,     
\end{equation}
%............
is rather insensitive to the effect of the redshift.

%%%%%%%%%%%%%%%%%%%%%%%%
\begin{table}[!t]
		\caption{Fitting parameters for the SN ALP spectrum, Eq.~\protect\ref{eq:time-int-spec1}, from the Primakoff process for different SN progenitors estimated for $g_{a\gamma}=10^{-11}\,\textrm{GeV}^{-1}$ and $m_a \ll 10^{-11}$~eV, assuming no gravitational energy-redshift/accounting for the gravitational redshift.
	 $\alpha$ particles contribution to the Primakoff production is included. \label{tab:spectfit_nored}
	 	}
		\begin{center}
			\begin{tabular}{lccc}
				\hline
				SN progenitor & $ $
				$C$ [$\times 10^{50} \,\  {\rm MeV}^{-1}$] \,\ 
				& $E_0$ [MeV] &$\beta$ \\
				\hline
				\hline
				8.8 $M_{\odot}$ & $4.18/4.81  \,\     $ & 90.62/78.15 &2.56/2.60 \\
				11.2 $M_{\odot}$ & $6.25/8.11  \,\     $ & 91.81/76.04 &2.74/2.80 \\
				18 $M_{\odot}$ & $18.4/26.1   \,\    $ & 119.4/89.32 &2.40/2.45 \\
				25 $M_{\odot}$ & $21.0/31.1   \,\    $ & 145.4/104.9 &2.25/2.30 \\
				40 $M_{\odot}$ & $1.56/2.98   \,\   $ & 168.9/110.6 &1.77/1.94 \\
				70 $M_{\odot}$ & $0.131/0.213  \,\     $ & 127.5/94.36 &1.13/1.76 \\
				\hline
			\end{tabular}
		\end{center}
	\end{table}
	%%%%%%%%%%%%%%%%%%%%%%%%%%%%%%%%
	
%		%%%%%%%%%%%%%%%%%%%%%%%%%%%%%%%
%	\begin{table}[!t]
%		\caption{Fitting parameters for the SN ALP spectrum, Eq.~(\eqref{eq:time-int-spec}), from the Primakoff process for different SN progenitors estimated for $g_{a\gamma}=10^{-11}\,\textrm{GeV}^{-1}$ and $m_a \ll 10^{-11}$~eV, accounting the gravitational energy-redshift.
%		}
%		\begin{center}
%			\begin{tabular}{lccc}
%				\hline
%				SN progenitor & $ $
%				$C$ [$\times 10^{50} \,\  {\rm MeV}^{-1}$] \,\ 
%				& $E_0$ [MeV] &$\beta$ \\
%				\hline
%				\hline
%				8.8 $M_{\odot}$ & $3.76  \,\     $ & 76.44 &2.59 \\
%				11.2 $M_{\odot}$ & $7.09  \,\     $ & 75.70 &2.80 \\
%				18 $M_{\odot}$ & $23.0   \,\    $ & 91.61 &2.43 \\
%				25 $M_{\odot}$ & $28.1   \,\    $ & 105.5 &2.30 \\
%				40 $M_{\odot}$ & $2.48   \,\   $ & 112.7 &1.92 \\
%				70 $M_{\odot}$ & $0.391  \,\     $ & 30.44 & 0.785 \\
%				\hline
%			\end{tabular}
%			\label{tab:spectfit_red}
%		\end{center}
%	\end{table}
%	%%%%%%%%%%%%%%%%%%%%%%%%%%%%%%%%

%%%%%%%%%%%%%%%%%%%%%%%	
		\begin{table}[!t]
		\caption{Fitting parameters for DSNALPB fluxes for  $g_{a\gamma} = 10^{-11}$ GeV$^{-1}$  and $m_a \ll 10^{-11}$~eV
			for different fractions of failed SNe $f_{\rm fail-CC}$, without gravitational energy-redshift/accounting for the gravitational redshift.  $\alpha$ particles contribution to the Primakoff production is included.
		}
		\begin{center}
			\begin{tabular}{lccc}
				\hline
				$f_{\textrm{fail-CC}}$ & $ $
					$C$  \,\ 
					& $E_0$  &$\beta$ \\
					& [$\times 10^{-7}  {\rm MeV}^{-1}\textrm{cm}^{-2}\textrm{s}^{-1}$] & [MeV] &\\
					\hline
					\hline
					10\%  \textrm{max flux}  & $96.2/144   \,\   $ & 58.5/43.8 &1.39/1.50 \\
					10\% & $58.0/88.9  \,\      $ & 59.3/43.5 &1.32/1.42 \\
					20\%  & $42.9/62.9  \,\     $ & 52.0/39.9 &1.41/1.49 \\
					30\%  & $31.0/46.5   \,\    $ & 50.8/39.3 &1.37/1.47 \\
					40\%  & $22.4/35.8   \,\    $ & 52.7/40.2 &1.28/1.41 \\
					40\% \textrm{min flux}  & $9.33/15.7  \,\     $ & 57.6/42.3 &1.17/1.32 \\
					\hline
				\end{tabular}
				\label{tab:dsnalp_nored}
			\end{center}
		\end{table}

%		%%%%%%%%%%
%			\begin{table}[!t]
%		\caption{Fitting parameters for DSNALPB fluxes for  $g_{a\gamma} = 10^{-11}$ GeV$^{-1}$  and $m_a \ll 10^{-11}$~eV
%			for different fractions of failed SNe $f_{\rm fail-CC}$ accounting for gravitational energy-redshift.
%		}
%		\begin{center}
%			\begin{tabular}{lccc}
%				\hline
%				$f_{\textrm{fail-CC}}$ & $ $
%					$C$ [$\times 10^{-7} \,\  {\rm MeV}^{-1}\textrm{cm}^{-2}\textrm{s}^{-1}$] \,\ 
%					& $E_0$ [MeV] &$\beta$ \\
%					\hline
%					\hline
%					10\%  \textrm{max flux}  & $210   \,\   $ & 42.6 &1.38 \\
%					10\% & $145  \,\     $ & 44.1 &1.39 \\
%					20\%  & $101  \,\     $ & 40.1 &1.46 \\
%					30\%  & $73.4   \,\    $ & 39.4 &1.43 \\
%					40\%  & $55.3   \,\    $ & 40.3 &1.37 \\
%					40\% \textrm{min flux}  & $28.0  \,\     $ & 41.6 &1.38 \\
%					\hline
%				\end{tabular}
%				\label{tab:dsnalp_red}
%			\end{center}
%		\end{table}	
%%%%%%%%%%%%

In order to quantify the impact of the gravitational energy-redshift on the  DSNALPB spectrum, 
 %and~\ref{tab:dsnalp_red}
we show the variation of the fitting parameters of  
Eq.~\ref{eq:time-int-spec1} in Table~\ref{tab:dsnalp_nored}, neglecting and including the gravitational energy-redshift effect, respectively.
We realize that the effect of gravitational energy-redshift  is the same 
observed on the spectrum of a single SN, i.e.~increase of the normalization factor $C$ and decrease in the average energy $E_0$.  The effect of the corrections on both parameters ranges between $\sim 25-35 \%$. We remark that the effect of the 
gravitational redshift is more sizable for failed SNe, which are never dominant in the DSNALPB flux, contributing at most at $40 \%$ of the SN progenitors. This would somehow dilute the final impact of the gravitational energy-redshift on the DSNALPB spectrum.\\

\newpage

\section{Systematic uncertainty on the DSNALPB upper limits of cosmological and astrophysical origin}
\label{app:B}

The following subsections contain a more detailed discussion of some of the sources of uncertainty regarding their impact on the ALP-photon coupling upper limits for the entirety of the relevant ALP mass range complementing the content of Tab.~\ref{tab:upperlim_uncertainty}.

\subsection{Impact of the fraction of failed CC SNe}
\label{sec:alp_vs_ffail}

As can be seen from the last column of Tab.~\ref{tab:upperlim_uncertainty}, the uncertainty of the fraction of failed CC SNe in the progenitor mass range chosen to compute the DSNALPB is the most important source of systematic uncertainty for this type of ALP-induced gamma-ray signal. 

\begin{figure*}[h!]
\begin{centering}
\includegraphics[width=0.75\linewidth]{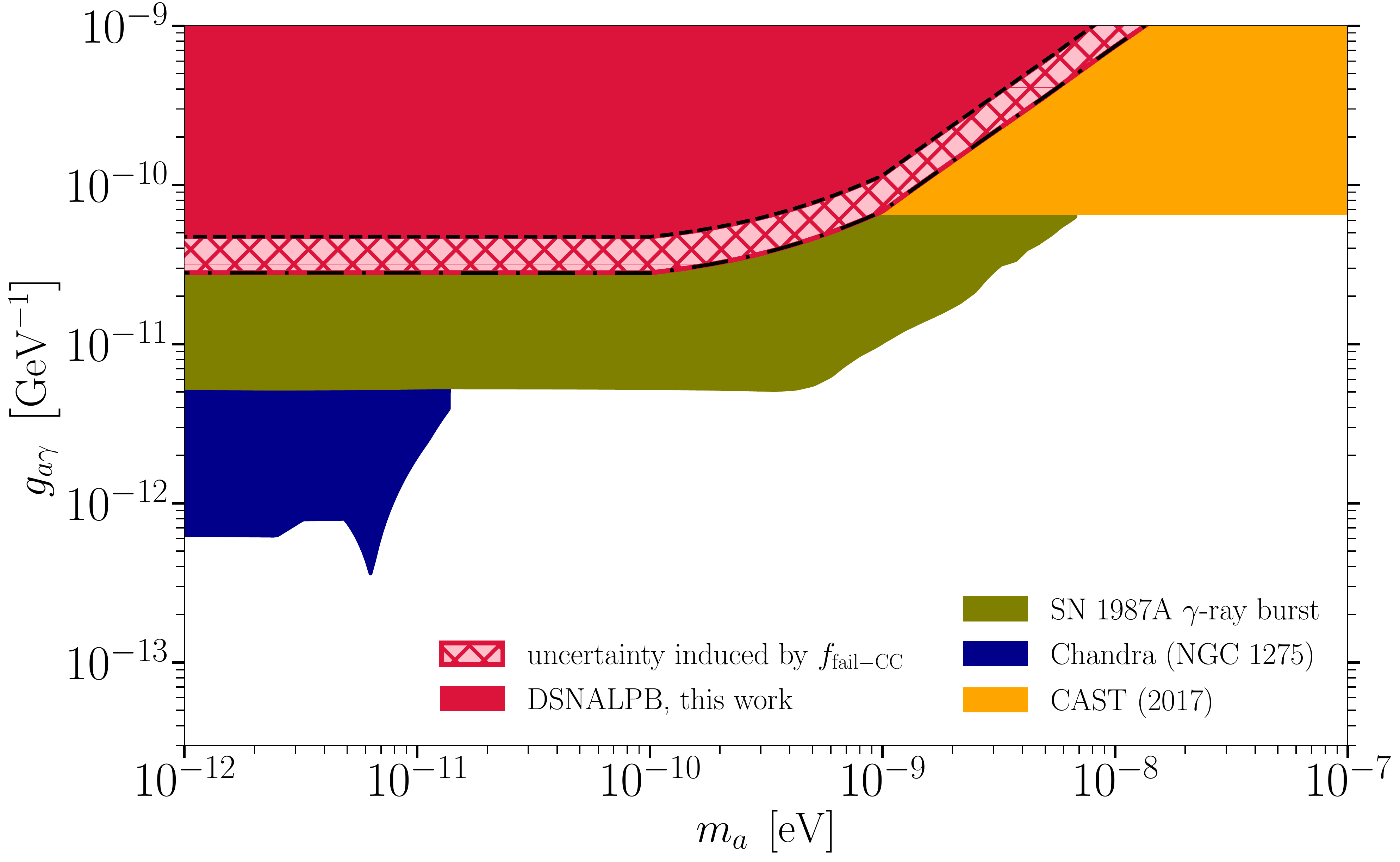}
\par\end{centering}
\caption{95$\%$ CL upper limits (red band) on the ALP-photon coupling constant $g_{a\gamma}$ assuming a coupling exclusively to photons and the `Jansson12c' \cite{2016A&A...596A.103P} model of the Milky Way's GMF. The displayed band reflects the uncertainty on the DSNALPB gamma-ray spectrum caused by the unknown ratio of failed to successful CCSNe within the mass range of SN progenitors considered in this analysis (see Sec.~\ref{sec:DSNALPB_theory}) while keeping all other properties as in the benchmark scenario. Our results are complemented by independent astrophysical and helioscope bounds on the ALP-photon coupling strength from CAST \cite{CAST:2017uph}, Chandra observations of NGC 1275 \cite{Reynolds:2019uqt} as well as the non-observation of a gamma-ray burst following SN 1987A \cite{Payez:2014xsa}.  \label{fig:limits_ALP_ffail}}
\end{figure*}

In Fig.~\ref{fig:limits_ALP_ffail}, we show the uncertainty band due to this source of systematic error in the full parameter space of ALPs.

\subsection{Impact of the Galactic magnetic field model}
\label{sec:alp_vs_bfield}

To assess the robustness of the upper limits presented in Sec.~\ref{sec:results} against different assumptions and models of the Milky Way's magnetic field, we create a sample of alternative signal templates which have been taken from a recent study of the \textsc{planck} collaboration \cite{2016A&A...596A.103P} (Tab.~3.1 therein) and \cite{2011ApJ...738..192P}.

\begin{figure*}[h!]
\begin{centering}
\includegraphics[width=0.75\linewidth]{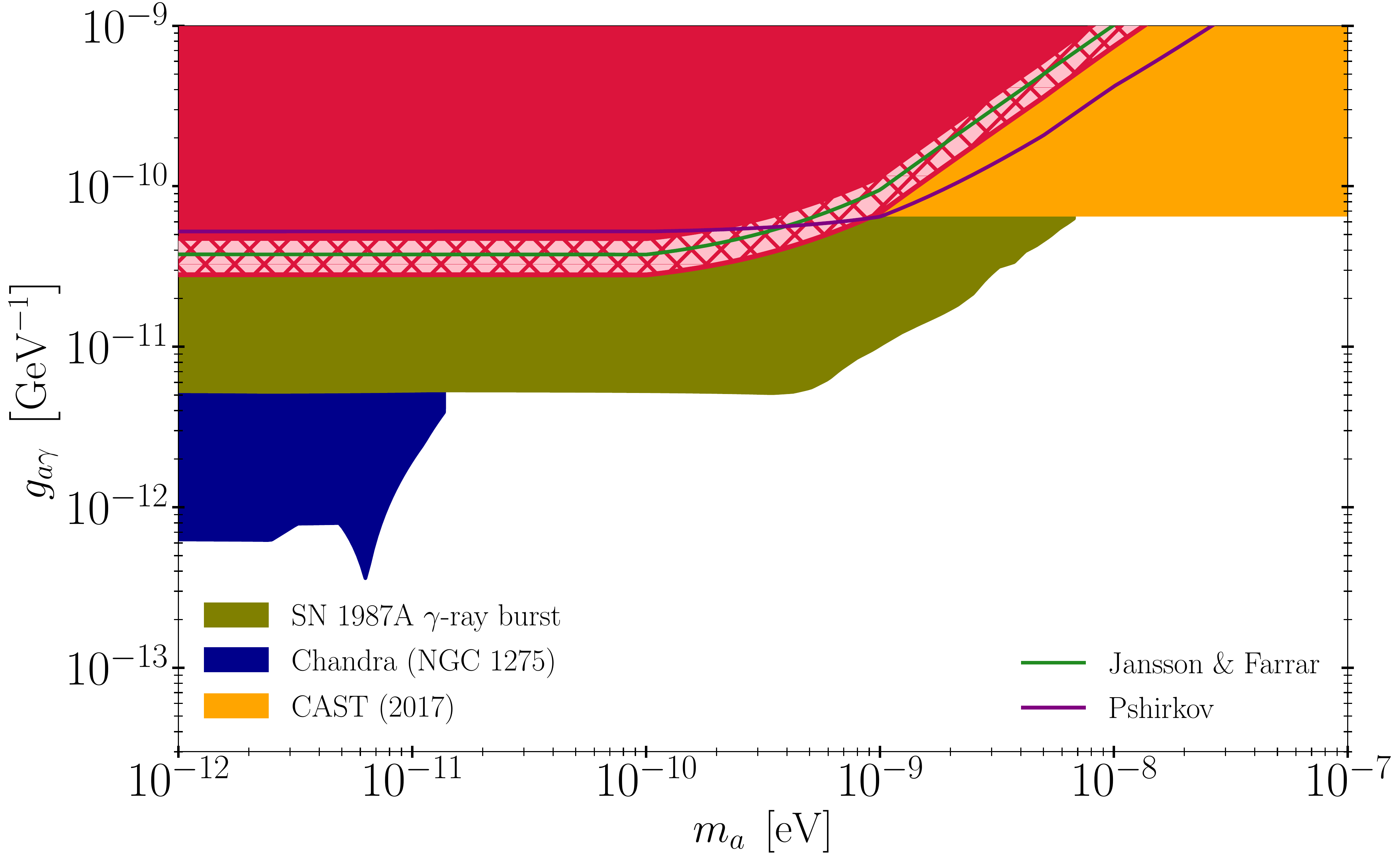}
\par\end{centering}
\caption{As in Fig.~\ref{fig:limits_ALP_ffail}. However, here we
focus on the variation with respect to the magnetic field model. 
We confront the upper limits derived with different characterisations of the Milky Way's magnetic field; `Jansson12c' \cite{2016A&A...596A.103P} (green) and `Pshirkov' \cite{2011ApJ...738..192P} (purple). For comparison, the theoretical uncertainty due to the fraction of failed and successful CC SNe is shown as a light red band. \label{fig:alp_vs_bfield}}
\end{figure*}

A comparison of the upper limits obtained from these models is displayed in Fig.~\ref{fig:alp_vs_bfield}. In general, different GMF models induce a variation in the derived upper limits on $g_{a\gamma}$ of $\mathcal{O}(1)$ whose relative impact on the final upper limit is comparable to the one of the $f_{\mathrm{fail-CC}}$ parameter according to Tab.~\ref{tab:upperlim_uncertainty}.

\subsection{Impact of the Galactic diffuse foreground model}
\label{sec:alp_vs_iem}

Although the ROI optimising has been conducted in the high-latitude gamma-ray sky to minimize the contamination by the Milky Way's diffuse foreground emission, we investigate the robustness of the upper limits shown in Sec.~\ref{sec:results} against variations of the Galactic foreground. To this end, we re-run the analysis pipeline with respect to the alternative IEMs introduced in Sec.~\ref{sec:methodology}. 

\begin{figure*}[h!]
\begin{centering}
\includegraphics[width=0.75\linewidth]{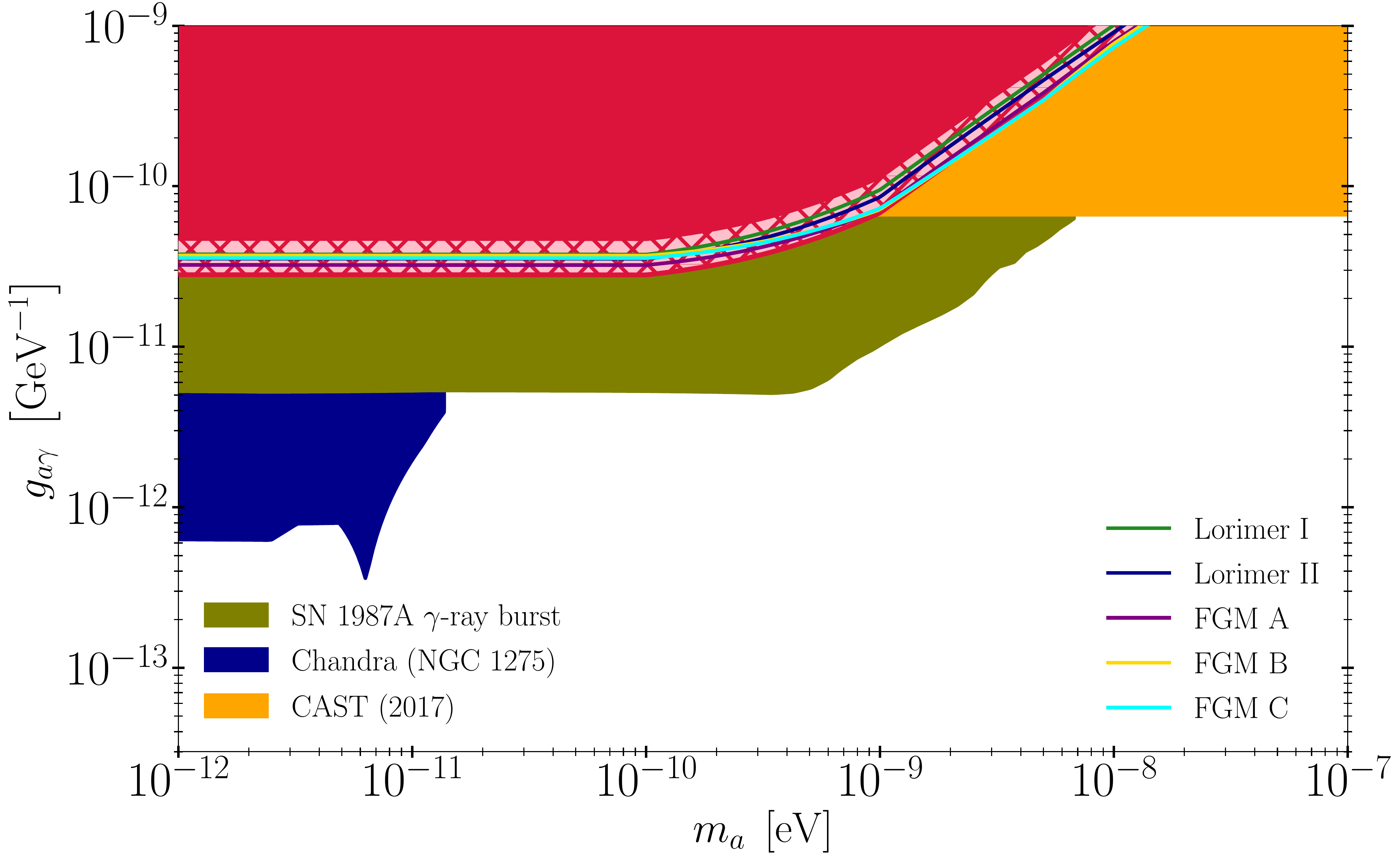}
\par\end{centering}
\caption{%Upper limits on the ALP-photon coupling constant $g_{a\gamma}$ assuming a coupling exclusively to photons. For definiteness, we present the upper limits with respect to the DSNALPB spectrum corresponding to $f_{\rm fail-CC} = 20\%$ (see Tab.~\ref{tab:fitting}). 
%($g_{aN} = 0$, solid) or an additional interaction term introducing a coupling to nucleons ($g_{aN} = 10^{-9}$, dashed).
As in Fig.~\ref{fig:limits_ALP_ffail} but here we focus on the dependence with respect to the background model. 
We compare the upper limits derived with our benchmark choice of IEM `Lorimer I' (green) with four alternative IEMs (c.f.~Sec.~\ref{sec:methodology}). Again, for comparison, the theoretical uncertainty due to the fraction of failed and successful CC SNe is shown as a light red band. \label{fig:alp_vs_iem}}
\end{figure*}

The results of this cross-check are presented in Fig.~\ref{fig:alp_vs_iem}. Variations of the IE have a smaller impact than model uncertainties in the magnetic field of the Milky Way. On one side, this implies that our analysis pipeline is robust against such alterations while, on the other side, it is essential to improve the existing models of the GMF, in particular, at high-latitudes.

\clearpage

\bibliographystyle{apsrev_mod}
\bibliography{alps_references.bib}

\end{document}